\documentclass[aps,prc,twocolumn ,superscriptaddress]{revtex4-1}
\usepackage{graphicx}
\usepackage{amsmath}
\usepackage{braket}
\usepackage{url}
\usepackage{ulem}
\usepackage{multirow}
\usepackage{amssymb} 
\usepackage{booktabs}
\usepackage{rotating}
\usepackage{bm}
\usepackage{bbm}
\usepackage{longtable}
\usepackage{subfigure}
\usepackage{tikz}
\usetikzlibrary{matrix}

\usepackage[colorlinks,
linkcolor=blue,
anchorcolor=red,
urlcolor=red,
citecolor=red]{hyperref}
\begin{document}


\title{Ground-state mass of $^{22}$Al and test of state-of-the-art \textit{ab initio} calculations}

\author{M.Z. Sun}
\affiliation{CAS Key Laboratory of High Precision Nuclear Spectroscopy, Institute of Modern Physics, Chinese Academy of Sciences, Lanzhou 730000, China}

\author{Y. Yu}
\affiliation{CAS Key Laboratory of High Precision Nuclear Spectroscopy, Institute of Modern Physics, Chinese Academy of Sciences, Lanzhou 730000, China}
\affiliation{School of Nuclear Science and Technology, University of Chinese Academy of Sciences, Beijing 100049, China}

\author{X.P. Wang}
\affiliation{CAS Key Laboratory of High Precision Nuclear Spectroscopy, Institute of Modern Physics, Chinese Academy of Sciences, Lanzhou 730000, China}
\affiliation{School of Nuclear Science and Technology, University of Chinese Academy of Sciences, Beijing 100049, China}  

\author{M. Wang}
\affiliation{CAS Key Laboratory of High Precision Nuclear Spectroscopy, Institute of Modern Physics, Chinese Academy of Sciences, Lanzhou 730000, China}
\affiliation{School of Nuclear Science and Technology, University of Chinese Academy of Sciences, Beijing 100049, China}

\author{J.G. Li}
\email{jianguo\_li@impcas.ac.cn}
\affiliation{CAS Key Laboratory of High Precision Nuclear Spectroscopy, Institute of Modern Physics, Chinese Academy of Sciences, Lanzhou 730000, China}

\author{Y.H. Zhang}
 \email{yhzhang@impcas.ac.cn}
\affiliation{CAS Key Laboratory of High Precision Nuclear Spectroscopy, Institute of Modern Physics, Chinese Academy of Sciences, Lanzhou 730000, China}
\affiliation{School of Nuclear Science and Technology, University of Chinese Academy of Sciences, Beijing 100049, China}

\author{K. Blaum}
\affiliation{Max-Planck-Institut f$\ddot{u}$r Kernphysik, Saupfercheckweg 1, 69117 Heidelberg, Germany}

\author{Z.Y. Chen}
\affiliation{CAS Key Laboratory of High Precision Nuclear Spectroscopy, Institute of Modern Physics, Chinese Academy of Sciences, Lanzhou 730000, China}
\affiliation{School of Nuclear Science and Technology, University of Chinese Academy of Sciences, Beijing 100049, China}

\author{R.J. Chen}
\affiliation{CAS Key Laboratory of High Precision Nuclear Spectroscopy, Institute of Modern Physics, Chinese Academy of Sciences, Lanzhou 730000, China}
\affiliation{GSI Helmholtzzentrum f$\ddot{u}$r Schwerionenforschung, Planckstraße 1, 64291 Darmstadt, Germany}

\author{H.Y. Deng}
\affiliation{CAS Key Laboratory of High Precision Nuclear Spectroscopy, Institute of Modern Physics, Chinese Academy of Sciences, Lanzhou 730000, China}
\affiliation{School of Nuclear Science and Technology, University of Chinese Academy of Sciences, Beijing 100049, China}

\author{C.Y. Fu}
\affiliation{CAS Key Laboratory of High Precision Nuclear Spectroscopy, Institute of Modern Physics, Chinese Academy of Sciences, Lanzhou 730000, China}

\author{W.W. Ge}
\affiliation{CAS Key Laboratory of High Precision Nuclear Spectroscopy, Institute of Modern Physics, Chinese Academy of Sciences, Lanzhou 730000, China}

\author{W.J. Huang}
\affiliation{Advanced Energy Science and Technology Guangdong Laboratory, Huizhou, 516007, China}
\affiliation{CAS Key Laboratory of High Precision Nuclear Spectroscopy, Institute of Modern Physics, Chinese Academy of Sciences, Lanzhou 730000, China}

\author{H.Y. Jiao}
\affiliation{CAS Key Laboratory of High Precision Nuclear Spectroscopy, Institute of Modern Physics, Chinese Academy of Sciences, Lanzhou 730000, China}
\affiliation{School of Nuclear Science and Technology, University of Chinese Academy of Sciences, Beijing 100049, China}

\author{H.H. Li}
\affiliation{CAS Key Laboratory of High Precision Nuclear Spectroscopy, Institute of Modern Physics, Chinese Academy of Sciences, Lanzhou 730000, China}
\affiliation{School of Nuclear Science and Technology, University of Chinese Academy of Sciences, Beijing 100049, China}

\author{H.F. Li}
\affiliation{CAS Key Laboratory of High Precision Nuclear Spectroscopy, Institute of Modern Physics, Chinese Academy of Sciences, Lanzhou 730000, China}

\author{Y.F. Luo}
\affiliation{CAS Key Laboratory of High Precision Nuclear Spectroscopy, Institute of Modern Physics, Chinese Academy of Sciences, Lanzhou 730000, China}
\affiliation{School of Nuclear Science and Technology, University of Chinese Academy of Sciences, Beijing 100049, China}

\author{T. Liao}
\affiliation{CAS Key Laboratory of High Precision Nuclear Spectroscopy, Institute of Modern Physics, Chinese Academy of Sciences, Lanzhou 730000, China}
\affiliation{School of Nuclear Science and Technology, University of Chinese Academy of Sciences, Beijing 100049, China}

\author{Yu.A. Litvinov}
\affiliation{CAS Key Laboratory of High Precision Nuclear Spectroscopy, Institute of Modern Physics, Chinese Academy of Sciences, Lanzhou 730000, China}
\affiliation{GSI Helmholtzzentrum f$\ddot{u}$r Schwerionenforschung, Planckstraße 1, 64291 Darmstadt, Germany}

\author{M. Si}
\affiliation{CAS Key Laboratory of High Precision Nuclear Spectroscopy, Institute of Modern Physics, Chinese Academy of Sciences, Lanzhou 730000, China}

\author{P. Shuai}
\affiliation{CAS Key Laboratory of High Precision Nuclear Spectroscopy, Institute of Modern Physics, Chinese Academy of Sciences, Lanzhou 730000, China}

\author{J.Y. Shi}
\affiliation{CAS Key Laboratory of High Precision Nuclear Spectroscopy, Institute of Modern Physics, Chinese Academy of Sciences, Lanzhou 730000, China}
\affiliation{School of Nuclear Science and Technology, University of Chinese Academy of Sciences, Beijing 100049, China}

\author{Q. Wang}
\affiliation{CAS Key Laboratory of High Precision Nuclear Spectroscopy, Institute of Modern Physics, Chinese Academy of Sciences, Lanzhou 730000, China}
\affiliation{School of Nuclear Science and Technology, University of Chinese Academy of Sciences, Beijing 100049, China}

\author{Y.M. Xing}
\affiliation{CAS Key Laboratory of High Precision Nuclear Spectroscopy, Institute of Modern Physics, Chinese Academy of Sciences, Lanzhou 730000, China}

\author{X. Xu}
\affiliation{CAS Key Laboratory of High Precision Nuclear Spectroscopy, Institute of Modern Physics, Chinese Academy of Sciences, Lanzhou 730000, China}

\author{H.S. Xu}
\affiliation{CAS Key Laboratory of High Precision Nuclear Spectroscopy, Institute of Modern Physics, Chinese Academy of Sciences, Lanzhou 730000, China}
\affiliation{School of Nuclear Science and Technology, University of Chinese Academy of Sciences, Beijing 100049, China}

\author{F.R. Xu}
\affiliation{State Key Laboratory of Nuclear Physics and Technology, School of Physics, Peking University, Beijing 100871, People’s Republic of China}

\author{Q. Yuan}
\affiliation{CAS Key Laboratory of High Precision Nuclear Spectroscopy, Institute of Modern Physics, Chinese Academy of Sciences, Lanzhou 730000, China}

\author{T. Yamaguchi}
\affiliation{Department of Physics, Saitama University, Saitama 338-8570, Japan}

\author{X.L. Yan}
\affiliation{CAS Key Laboratory of High Precision Nuclear Spectroscopy, Institute of Modern Physics, Chinese Academy of Sciences, Lanzhou 730000, China}

\author{J.C. Yang}
\affiliation{CAS Key Laboratory of High Precision Nuclear Spectroscopy, Institute of Modern Physics, Chinese Academy of Sciences, Lanzhou 730000, China}
\affiliation{School of Nuclear Science and Technology, University of Chinese Academy of Sciences, Beijing 100049, China}

\author{Y.J. Yuan}
\affiliation{CAS Key Laboratory of High Precision Nuclear Spectroscopy, Institute of Modern Physics, Chinese Academy of Sciences, Lanzhou 730000, China}
\affiliation{School of Nuclear Science and Technology, University of Chinese Academy of Sciences, Beijing 100049, China}

\author{X.H. Zhou}
\affiliation{CAS Key Laboratory of High Precision Nuclear Spectroscopy, Institute of Modern Physics, Chinese Academy of Sciences, Lanzhou 730000, China}
\affiliation{School of Nuclear Science and Technology, University of Chinese Academy of Sciences, Beijing 100049, China}

\author{X. Zhou}
\affiliation{CAS Key Laboratory of High Precision Nuclear Spectroscopy, Institute of Modern Physics, Chinese Academy of Sciences, Lanzhou 730000, China}

\author{M. Zhang}
\affiliation{CAS Key Laboratory of High Precision Nuclear Spectroscopy, Institute of Modern Physics, Chinese Academy of Sciences, Lanzhou 730000, China}

\author{Q. Zeng}
\affiliation{CAS Key Laboratory of High Precision Nuclear Spectroscopy, Institute of Modern Physics, Chinese Academy of Sciences, Lanzhou 730000, China}
\affiliation{School of Nuclear Science and Engineering, East China University of Technology, Nanchang 330013, China}



\date{\today}

\begin{abstract}
The ground-state mass excess of the $T_{z}=-2$ drip-line nucleus $^{22}$Al is measured for the first time to be $18103(10)$ keV using the newly-developed B$\rho$-defined isochronous mass spectrometry method at the cooler storage ring in Lanzhou. The new mass excess value allowed us to determine the excitation energies of the two low-lying $1^+$ states in $^{22}$Al with significantly reduced uncertainties of 51 keV. Comparing to the analogue states in its mirror nucleus $^{22}$F, the mirror energy differences of the two $1^+$ states in the $^{22}$Al-$^{22}$F mirror pair are determined to be $-625(51)$ keV and $-330(51)$ keV, respectively. The excitation energies and the mirror energy differences are used to test the state-of-the-art \textit{ab initio} valence-space in-medium similarity renormalization group calculations with four sets of interactions derived from the chiral effective field theory. The mechanism leading to the large mirror energy differences is investigated and attributed to the occupation of the $\pi s_{1/2}$ orbital.
\end{abstract}



\maketitle


\section{Introduction}

In recent years, \textit{ab initio} calculations have made significant progress, largely due to the introduction of interactions issued from chiral effective field theory~\cite{HERGERT2016165,doi:10.1146/annurev-nucl-101917-021120,3,5}. Among these achievements, the valence-space in-medium similarity renormalization group approach (VS-IMSRG) stands out. This novel approach has dramatically expanded the scope of \textit{ab initio} calculations, bridging the gap from primarily light-mass region to encompassing medium-mass domain~\cite{4,5,31,33,wang_mass_2023,PhysRevLett.126.022501}. 
A pivotal insight from these advanced calculations is the crucial role of the three-nucleon force (3NF). The inclusion of 3NF has proven instrumental in deepening our understanding of nuclear properties and elucidating the intricate architecture of nuclear structure~\cite{5,6,PhysRevLett.126.022501,wang_mass_2023}. Nowadays, \textit{ab initio} calculations have demonstrated themselves as a robust tool in nuclear physics, offering powerful adaptability for addressing fundamental questions in nuclear structure physics~\cite{4,5}.

Up to now, the application of \textit{ab initio} calculations in nuclear structure theory has been mainly concentrated on even-even and odd-A nuclei. This preference arises because their structure is predominantly influenced by single-particle motions and pairing correlations, which are well addressed by \textit{ab initio} calculations. Indeed, the results of \textit{ab initio} calculations for structure of even-even and odd-A nuclei align commendably with experimental data~\cite{PhysRevC.107.014302,zhang_roles_2022,LI2023138197}. However, when considering odd-odd nuclei, establishing a correspondence between the experimental and calculated states becomes challenging. The proton-neutron interaction between unpaired protons and neutrons complicates the matter further. Consequently, the odd-odd nucleus presents a complementary and stringent testing ground for the \textit{ab initio} approaches~\cite{5}, especially for the drip-line nuclei. Currently, only a few applications of \textit{ab initio} calculations have been performed for the odd-odd nuclei~\cite{hu_ab-initio_2020,PhysRevLett.110.082502,PhysRevC.96.054305}.





The level structure of mirror nuclei are commonly addressed and discussed on the basis of isospin symmetry, which is a basic assumption in particle and nuclear physics. However, this symmetry is known to be approximate, and the corresponding deviation is called isospin symmetry breaking (ISB)~\cite{PhysRevLett.89.142502, PhysRevLett.92.132502, PhysRevLett.97.132501, PhysRevLett.97.152501, PhysRevLett.110.172505}. Studies of mirror nuclei offer profound insights into the origin of ISB and further information about nuclear structure, as well as facilitating evaluations of nuclear models~\cite{PhysRevLett.89.142502, PhysRevLett.92.132502, PhysRevLett.97.132501, PhysRevLett.97.152501, PhysRevLett.110.172505,PhysRevLett.125.192503,PhysRevC.107.014302,CPC:10.1088/1674-1137/acf035}. One of the key quantities in the investigation of ISB in mirror nuclei is the mirror energy difference (MED)~\cite{PhysRevLett.89.142502,2006ARNPS..56..253M,LI2023138197}, which is defined as the difference of excitation energies between the analogue states in mirror nuclei. The values of MED amount to a few tens or hundreds of keV~\cite{PhysRevLett.89.142502,2006ARNPS..56..253M}. A proton-rich nucleus close to the drip line can exhibit characteristics of weakly-bound unbound system in contrast to the deeply-bound system of  its neutron-rich partner.
Such disparities lead to pronounced MED values of mirror nuclei, exhibiting significant ISB. This phenomenon is known as Thomas-Ehrman shift (TES)~\cite{PhysRev.88.1109, PhysRev.81.412}.

$^{22}$Al is the lightest bound Al isotope with $\mathrm{T_{z}=(N-Z)/2=-2}$. Two low-lying $1^+$ states in odd-odd $^{22}$Al have been identified via the  $\beta$-delayed one-proton emissions from $^{22}$Si~\cite{PhysRevLett.125.192503}.  However, their excitation energies have not been precisely determined due to the lack of the experimental ground-state mass of $^{22}$Al. Using the extrapolated mass value of $^{22}$Al in Atomic Mass Evaluation 2020 (AME$\mathrm{'}$20)~\cite{18} and the analogue states of its mirror partner $^{22}$F~\cite{endsf}, significant isospin symmetry breaking was observed with large MED values of $-722(403)$ keV and $-427(403)$ keV for the $1_{1,2}^+$ states of $^{22}$Al-$^{22}$F mirror pair, respectively. 
A large isospin asymmetry in the $^{22}$Si-$^{22}$O mirror Gamow-Teller transitions to the $\mathrm{1^{+}_{1}}$ states of their daughter nuclei ($^{22}$Al-$^{22}$F) was observed and explained via the shell-model calculations as due to the loosely bound $s_{1/2}$ proton in $^{22}$Al, which could be  the origin of possible halo structure for the $1^+_1$ state~\cite{PhysRevLett.125.192503}.
 \textit{Ab initio} calculations have also been applied to investigate the large MEDs of the sd-shell nuclei~\cite{PhysRevC.107.014302,CPC:10.1088/1674-1137/acf035,PhysRevC.108.L031301}. The significant uncertainties in MEDs of $^{22}$Al-$^{22}$F mirror nuclei hinder direct comparison with theoretical predictions, thereby challenging our understanding of the isospin symmetry breaking in the dripline nucleus $^{22}$Al. Thus, precision mass measurement of the ground state of $^{22}$Al is urgently needed.

In this work, we report the first mass measurement of the ground state of $^{22}$Al, which enables us to determine the MEDs of the $^{22}$Al-$^{22}$F mirror pair. The large MEDs are investigated using the \textit{ab initio} calculations with several sets of interactions derived from chiral effective field theory, providing a stringent and complementary testing ground for the state-of-the-art theoretical model.

\section{Experiment and data analysis}

\begin{figure}
\centering
\includegraphics[width=0.9\linewidth]{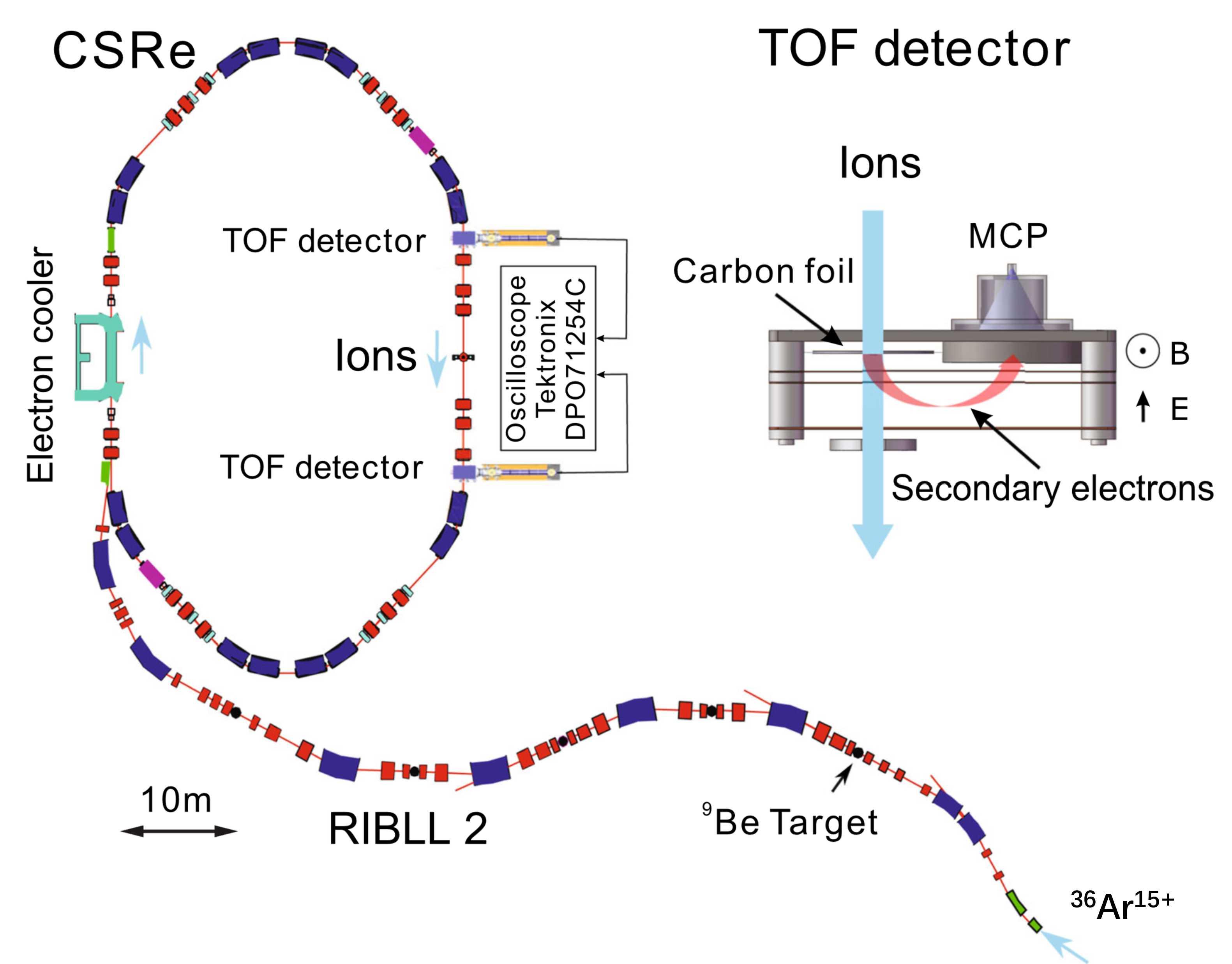}
\caption{Layout of RIBLL2 and CSRe machines at IMP, Lanzhou. A pair of identical TOF detectors are installed in the straight section of CSRe.}\label{CSR}
\end{figure}

\begin{figure}
\centering
\includegraphics[width=1.0\linewidth]{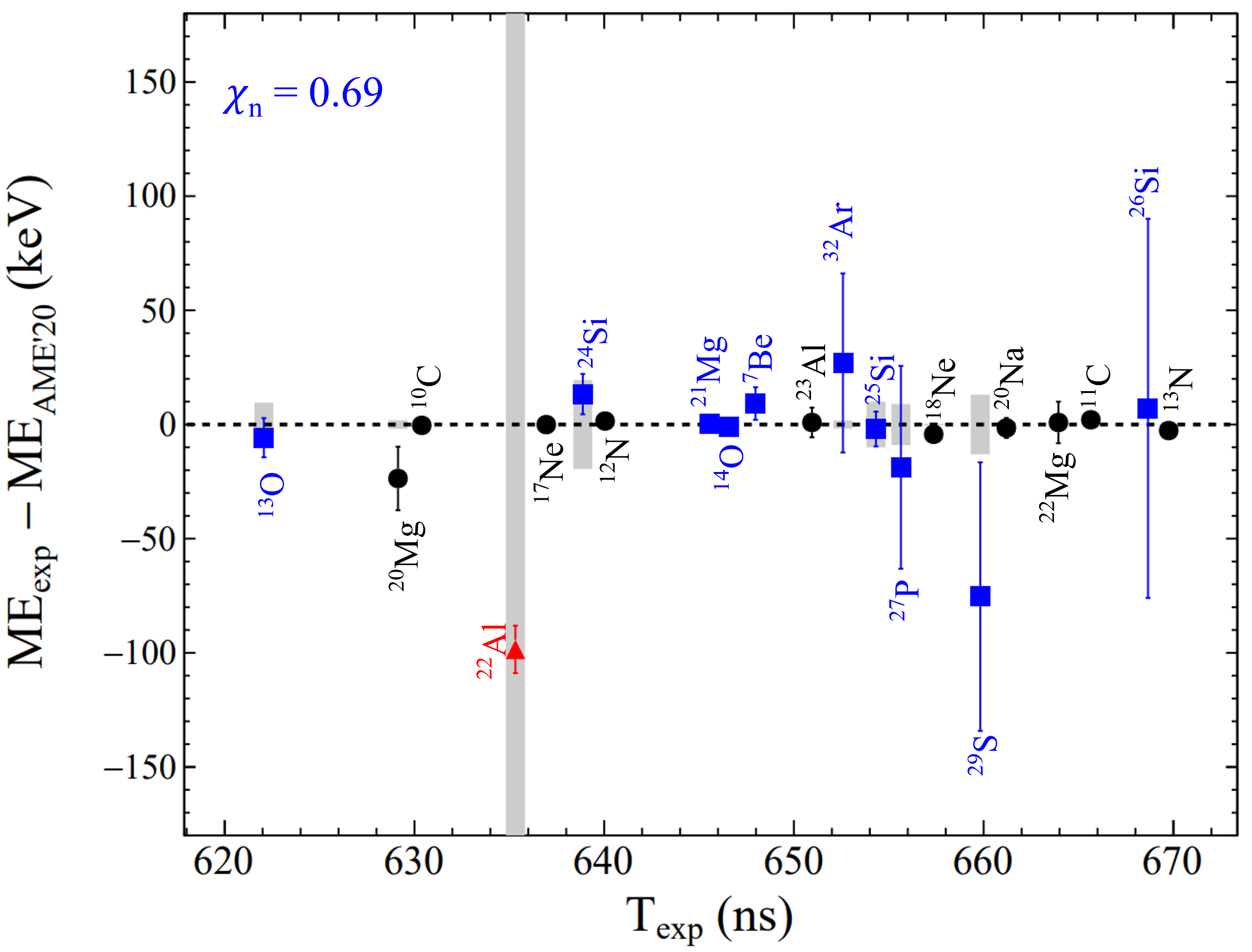}
\caption{Comparison of mass excess (ME) values determined in this work with the literature values in AME$\mathrm{'}$20~\cite{18}. The black circles represent the nuclides used as calibrants, the blue squares stand for the ones used for checking the reliability of our measurement. The mass uncertainties in AME$\mathrm{'}$20 are represented by the gray shadows. The mass of $^{22}$Al (red triangle) is measured for the first time in this work.}\label{MECali}
\end{figure}

The ground-state mass of $^{22}$Al was measured by using the newly-established B$\mathrm{\rho}$-defined IMS technique~\cite{17,zhang_brho_2023}. The experiment was conducted at the HIRFL-CSR accelerator complex at the Institute of Modern Physics (IMP) in Lanzhou, which is composed of the heavy-ion synchrotron CSRm, the in-flight fragment separator RIBLL2, and the experimental cooler storage ring CSRe~\cite{9,10} (see Fig.~\ref{CSR}). In this experiment, beams of $\mathrm{^{36}Ar^{15+}}$ were accumulated and accelerated to a relativistic energy of 400.505 MeV/u by CSRm. Every 25 s, the high-energy beam was fast extracted and focused upon a 15 mm beryllium target placed at the entrance of RIBLL2. The reaction fragments, which were fully stripped, were separated in flight with RIBLL2 and a cocktail beam was injected into CSRe. The RIBLL2-CSRe system was set to a fixed central magnetic rigidity of 4.841 Tm. CSRe was characterized by a maximum momentum acceptance of $\mathrm{\pm0.33\%}$, and tuned to an isochronous ion-optical mode with the transition point $\mathrm{\gamma_{t}=1.339}$~\cite{tu_direct_2011,12}. On average, about nine ions were stored simultaneously in each injection.

The passing time of the ion was measured by a pair of identical TOF detectors~\cite{15} installed in a straight section of CSRe~\cite{14} (see Fig.~\ref{CSR}). Each detector consists of a thin carbon foil ($\mathrm{\phi}$40 mm, thickness 18 $\mathrm{ \mu g/cm^{2}}$) and a set of micro-channel plates (MCP). When an ion passed through the carbon foil, secondary electrons were released from the foil surface and guided to the MCP by perpendicularly arranged electric and magnetic fields. Fast timing signals from the anode of the detector were recorded by a digital oscilloscope at a sampling rate of 50 GHz. The measurement time was 400 $\mathrm{\mu s}$ long for each injection, where the recorded two trains of timing signals corresponded to $\sim$630 revolutions of the ions in the ring.

A standard data analysis method for B$\mathrm{\rho}$-defined IMS was employed~\cite{17,zhang_brho_2023}. The passing times of the ions were extracted from the recorded signals using the constant fractional discrimination technique. From the two timing sequences of each ion, the revolution time $T$ as well as its velocity $v$ was extracted as described in Ref.~\cite{16}.
Particle identifications were made by comparing the obtained experimental revolution time spectrum with the simulated one~\cite{tu_nima_2011}. The magnetic rigidity $B\rho$ and orbit length $C$ were thus determined according to
\begin{equation}
B\rho = \frac{m}{q} v \gamma \text{,   ~and~~~ ~}  C = T v.
\end{equation}
The \{$B\rho_{exp}$, $C_{exp}$\} dataset for the nuclides with well-known masses (mass uncertainty smaller than 5 keV in literature~\cite{18}) and with more than 100 recorded events were used to construct the $B\rho(C)$ function. Then, the $m/q$ value of any ion including the unknown-mass nuclei was derived directly according to 
\begin{equation}
(\frac{m}{q})_{exp} = \frac{B\rho(C_{exp})}{(\gamma v)_{exp}}.
\end{equation}
Ten nuclides (black circles in Fig. 2) were used for calibration, i.e., to construct the $B\rho(C)$ function. To check the reliability of our measurement, masses of other ten known-mass nuclei (blue squares in Fig. 2) were redetermined which agree well with literature values. Their normalized chi-square $\chi_{n}$=0.69 indicated that no additional systematic errors were needed. 
$^{7}$Be and $^{32}$Ar were not used as calibrants due to their low statistics. Neither $^{21}$Mg nor $^{14}$O were used as calibrants because they could not be completely separated in the revolution time spectrum.
Their mass values were obtained by performing double-Gaussian fitting on the $m/q$ spectrum.

\begin{figure}
\centering
\includegraphics[width=1.0\linewidth]{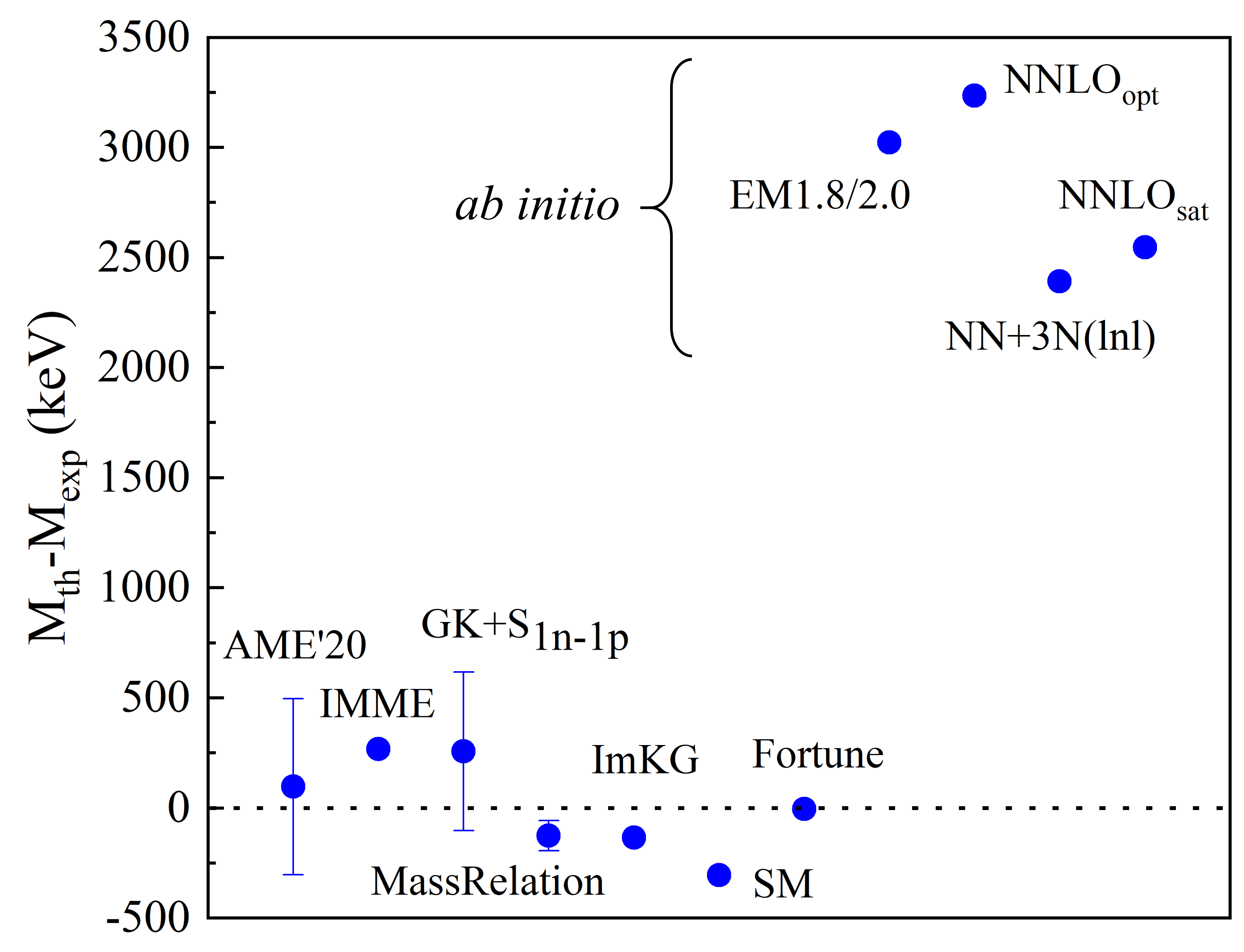}
\caption{
 Comparison of the $^{22}$Al ground-state masses predicted by various mass models with the experimental value reported in this work. The predicted mass values are drawn with error bars except for the ones labled as ``SM", ``Fortune" and ``\textit{ab initio}''.}  \label{mass-22Al}
\end{figure}

\section{Results and discussion}

The ground-state mass excess of $^{22}$Al was determined to be ME=$18103(10)$ keV, which is 97(400) keV smaller than the extrapolated ME=$18200(400)$ keV in AME$\mathrm{'}$20~\cite{18}. The proton separation energy of $^{22}$Al was then obtained to be $S_{p}$=90(10) keV, showing that this nucleus is weakly bound. The measured ground-state mass of $^{22}$Al is compared with predictions from various mass models  in Fig.~3 
including AME$\mathrm{'}$20~\cite{18}, isobaric multiplet mass equation (IMME)~\cite{kondev_nubase2020_2021}, 
improved mass extrapolations from the Garvey-Kelson relations (GK+S$\mathrm{_{1n-1p}}$)~\cite{24}, mass relations of mirror nuclei (MassRelation)~\cite{25}, improved Kelson-Garvey mass relations (ImKG)~\cite{26}, shell model calculation (SM)~\cite{27}, mass prediction via parametrization of the differences between separation energies for neutrons and protons in mirror nuclei  (Fortune)~\cite{28} and \textit{ab initio} calculations (\textit{ab initio})~\cite{PhysRevLett.126.022501}. 

Among all the predicted masses, the value given by Fortune aligns most closely with the measured one.  The nearly 250-keV difference from the IMME  prediction is most likely due to  the misidentification of the T = 2, J$^\pi$ = 4$^{+}$ isobaric analog state in $^{22}$Ne~\cite{kondev_nubase2020_2021,MACCORMICK201461}. The mass values predicted by the present \textit{ab intio} calculations deviate about 3 MeV from the experimental value. In fact, about 3.3 MeV deviation was already found in the earlier \textit{ab intio} calculations for the nuclei from helium through iron~\cite{PhysRevLett.126.022501}. Such systematic deviation in the absolute binding energies could be washed out when addressing the excitation energies~\cite{PhysRevC.107.014302,PhysRevC.108.L031301,doi:10.1146/annurev-nucl-101917-021120}.

\begin{table*}
\caption{\label{tab:table1}The excitation energies of the two $1^+$ states of $^{22}$Al and their corresponding MEDs. The ground-state mass of $^{22}$Al reported in this work was used in columns 2 and 3, while the mass in AME$\mathrm{'}$20~\cite{18} was used in columns 4 and 5. All energies are in keV.}
\begin{ruledtabular}
\begin{tabular}{ccccc}

 $\mathrm{J^{\pi}_{i}}$& $E_x$(IMS)& MED(IMS)& $E_x$(AME$\mathrm{'}$20)& MED(AME$\mathrm{'}$20)\\ \hline
 $\mathrm{1^{+}_{1}}$ & 1002(51)& $-$625(51)& 905(403)& $-$722(403)\\
 $\mathrm{1^{+}_{2}}$ & 2242(51)&$-$330(51)& 2145(403)& $-$427(403)\\
\end{tabular}
\end{ruledtabular}
\end{table*}

Thomas-Ehrman shift occurs in the proton drip-line nuclei, where a large MED is exhibited. It provides a unique laboratory to investigate the structural intricacies of nuclei with imbalanced neutron-to-proton ratios~\cite{bentley_coulomb_2007,boso_neutron_2018,PhysRevLett.125.192503,27,PhysRevC.107.014302,LI2023138197}. The MED is expressed as 
\begin{equation}
\text{MED}=E_{x}(J,T,T_{z}=-T)-E_{x}(J,T,T_{z}=T),
\end{equation}
where $E_{x}(J,T,T_{z})$ denotes the excitation energy of a state with spin $J$, isospin $T$ and z-projection  $T_{z}$. With the precisely measured ground-state mass of $^{22}$Al, the excitation energies of the two $1^+$ states of $^{22}$Al are obtained by using the reported data of $\beta$-delayed proton emissions from $^{22}$Si~\cite{PhysRevLett.125.192503} and the known ground-state mass of $^{21}$Mg~\cite{18}. Comparing to the analogue states in its mirror nucleus $^{22}$F~\cite{endsf}, the MEDs of the two $1^+$ states in $^{22}$Al-$^{22}$F mirror partners are extracted and presented in Table I, along with the previous results using the mass of $^{22}$Al in AME$\mathrm{'}$20~\cite{18}. It is obvious that the absolute values of the MEDs for both states are reduced by 97 keV relative to the previous ones. Moreover, the associated uncertainties have been notably reduced from 403 keV to 51 keV originating mainly from the energy uncertainties of the $\beta$-delayed protons from $^{22}$Si~\cite{PhysRevLett.125.192503}.

The MEDs of $^{22}$Al-$^{22}$F odd-odd mirror nuclei were calculated by an \textit{ab initio} valence-space in-medium similarity renormalization group approach (VS-IMSRG), employing several sets of nuclear forces derived from chiral effective field theory. The VS-IMSRG approach constructs a unitary transformation from the Magnus formalism \cite{HERGERT2016165,doi:10.1146/annurev-nucl-101917-021120}. The effective Hamiltonian within the full $sd$-shell for valence protons and neutrons above the $^{16}$O inner core is constructed. The ensemble normal-ordering procedure is adopted to further capture the effects of 3N forces~\cite{4,imsrg_code}. All operators are truncated at the normal-ordered two-body level during the VS-IMSRG calculations, whereby the VS-IMSRG code of Ref.~\cite{imsrg_code} is utilized for that matter. Finally, the derived effective Hamiltonian is exactly diagonalized using the shell-model code from Ref.~\cite{39}, facilitating our exploration of the physics of interest in the present work and allowing us to test nuclear forces in any fully open-shell system accessible to the nuclear shell model.

\begin{figure*}
\centering
\includegraphics[width=1.0\linewidth]{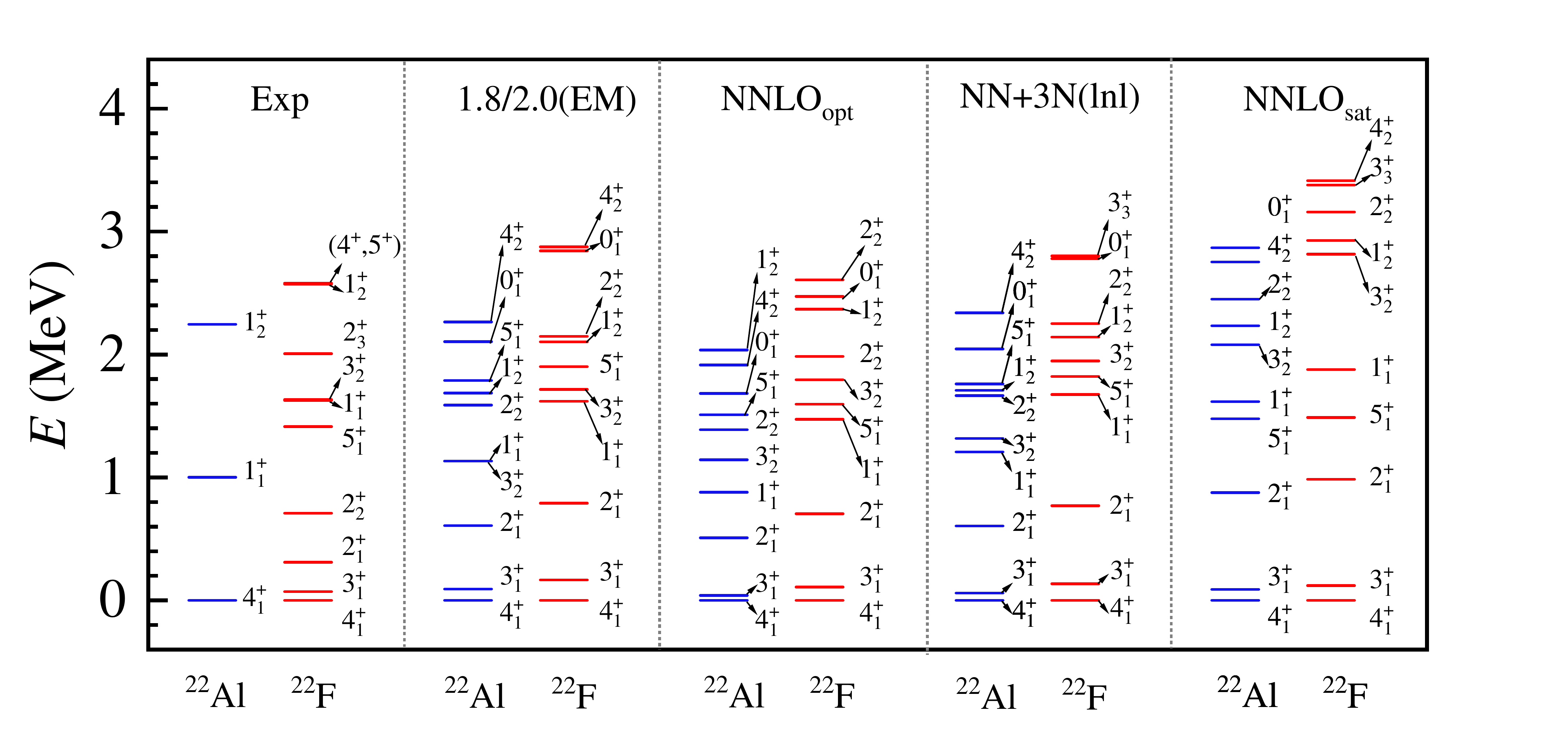}
\caption{The calculated spectra of $^{22}$Al/$^{22}$F mirror nuclei using \textit{ab initio} VS-IMSRG calculations with four sets of nuclear interactions, including 1.8/2.0 (EM), $NN+3N(\text{lnl})$, and NNLO$_{\text{sat}}$ $NN +3N$ interactions,  and NNLO$_{\text{opt}}$ $NN$ interaction,  along with the experimental spectrum.}\label{fig4} 
\end{figure*}

In this work, we employ several sets of chiral interactions. Firstly, 1.8/2.0(EM) is made up of the initial chiral next-to-next-to-next-to-leading order (N$^3$LO) \textit{NN} force \cite{PhysRevC.68.041001}  softened by a similarity renormalization group (SRG) evolution \cite{PhysRevC.75.061001} using $\lambda_{\rm SRG} = 1.8$ fm$^{-1}$, and next-to-next-to-leading order (N$^2$LO) $3N$ interaction with a non-locally regulation in which the cutoff $\Lambda = 2.0$ fm$^{-1}$. 
The adopted $NN$ interaction provides accurate descriptions of the nucleon-nucleon scattering phase shift, and the low-energy constants of three-body forces are fitted in $A = 3$ and 4-body systems. The $NN + 3N$ EM1.8/2.0 potential successfully reproduces ground- and excited-state energies to the tin region and beyond~\cite{31,wang_mass_2023,PhysRevLett.126.022501,PhysRevC.105.014302}.
Secondly, we utilize a newly developed chiral potential at N$^3$LO combined with an N$^2$LO $3N$ interaction including both local and nonlocal $(\text{lnl}) 3N$ regulators, labeled by $NN+3N(\text{lnl})$ \cite{PhysRevC.101.014318}. This interaction provides a good description of ground-state energies from oxygen to nickel isotopes.  The induced three-body force is neglected in the present work by employing a large similarity renormalization group cutoff of $\lambda = 2.6$ fm$^{-1}$ for $NN+3N(\text{lnl})$ interaction.
Next, N$^2$LO$_{\text{sat}}$ is adopted, which is constructed from the $NN$ and $3N$ both up to the chiral N$^2$LO, and fit to medium-mass data. The N$^2$LO$_{\text{sat}}$ interaction reproduces ground-state energies and radii to the nickel region \cite{33,34,3,36,37}.  Lastly, we adopt NNLO$_{\text{opt}}$, an interaction constructed from the chiral N$^2$LO $NN$, which yields good agreement for the spectra and position of the neutron drip line in oxygen, shell-closures in calcium, and the neutron matter equation of state at subsaturation densities, without resorting to $3N$ forces \cite{PhysRevLett.110.192502}.
In practical calculations, the harmonic oscillator (HO) basis is used to define the model space. Specifically, we use 15 HO major shells (i.e.~$e=2n+l \leq e_{\mathrm{max}}=14$) for $NN$ interaction, and additional truncation on the $3N$ matrix elements is limited as well to $e_{3\text{max}} = 2n_a+2n_b+2n_c+l_a+l_b+l_c\leq 14$.  The Coulomb force is also included in the calculations.

\begin{figure}
\centering
\includegraphics[width=1.0\linewidth]{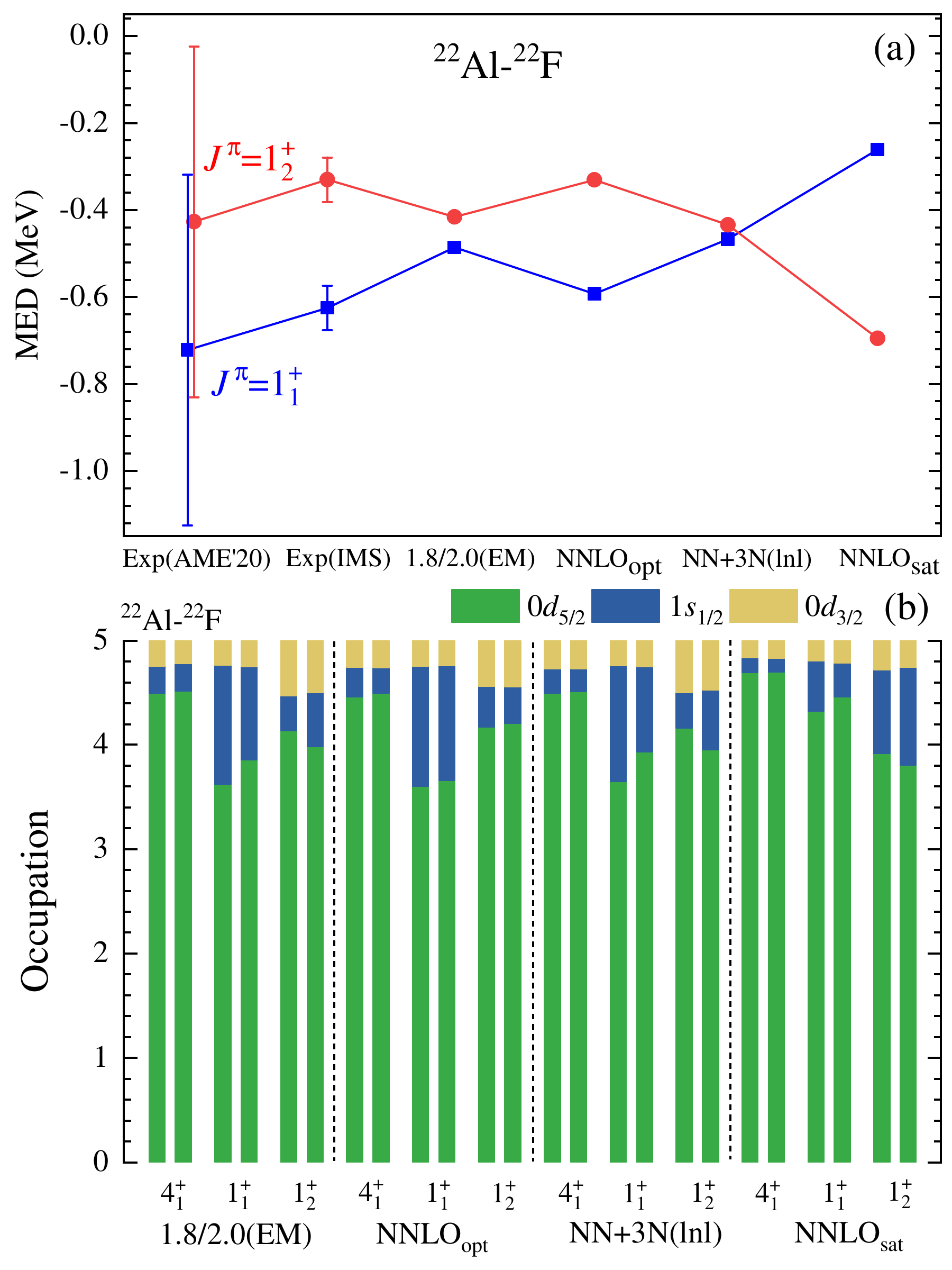} 
\caption{Mirror energy differences (a) and calculated average occupations of single-particle valence orbits (b) of the $\mathrm{1^{+}_{1,2}}$ states in $^{22}$Al/$^{22}$F mirror nuclei by employing \textit{ab initio} VS-IMSRG calculations with four sets of nuclear interactions. The experimental MED values are drawn with error bars while the calculated values are not. The average occupations correspond to the valence protons and neutrons in $^{22}$Al and $^{22}$F, respectively.}\label{fig5}
\end{figure}

The ground-state mass of $^{22}$Al predicted by \textit{ab inito} VS-IMSRG calculations using the four chiral potentials are displayed in Fig.~\ref{mass-22Al}. The calculated low-lying spectra of $^{22}$Al-$^{22}$F are presented in Fig.~\ref{fig4} along with available experimental data. One see that  the calculations agree well with experiment within 1 MeV except for the $2_2^+$ state.  In particular, the level ordering of the lowest $4_1^+$, $3_1^+$, and $2_1^+$ states in $^{22}$F are correctly reproduced.
A consistent result across all calculations is that the excitation energies of the two $1^+$ state in $^{22}$Al are invariably lower than its counterparts in  $^{22}$F, aligning well with experimental data. From the above results,  although there are still disparities in the ordering for some states comparing with the experimental data, the \textit{ab initio} calculations furnish commendable representations of the low-lying states in the $^{22}$Al-$^{22}$F mirror nuclei.




To further test the \textit{ab initio} calculations with different interactions, MEDs of the $1_{1,2}^+$ states in the $^{22}$Al-$^{22}$F mirror nuclei were calculated and shown in Fig. \ref{fig5}(a) along with the experimental values obtained in this work and the results from Ref.~\cite{PhysRevLett.125.192503}.  
The large uncertainties of the MEDs obtained in Ref.~\cite{PhysRevLett.125.192503} prevent us from benchmarking theoretical calculations. The uncertainties of MEDs obtained in this work are significantly reduced, and the results show that the MED of the $1_1^+$ states is larger than that of the $1_2^+$ states in $^{22}$Al-$^{22}$F mirror nuclei. 
From the \textit{ab initio} calculations, we can see that the MEDs calculated with  1.8/2.0(EM), NNLO$_{\text{opt}}$, and $NN+3N(\text{lnl})$ interactions agree with the experimental data, especially the result from NNLO$_{\text{opt}}$, 
while the predicted MED values with NNLO$_{\text{sat}}$  for the $1^+_{1,2}$ states are inverted comparing to the experimental data. Therefore, the MED of mirror nuclei serves as a more sensitive observable for testing theoretical calculations when comparing the excitation spectra.
Moreover, the mirror systems $^{22}$Al-$^{22}$F have been also investigated using \textit{ab initio} Gamow shell model based on the same EM1.8/2.0 interaction~\cite{PhysRevC.108.L031301}. Within the Gamow shell model calculations, the effective Hamiltonian is derived through the many-body perturbation theory, and continuum coupling is well accounted for using the Berggren basis. 
The results obtained from \textit{ab initio} GSM calculations are consistent with those obtained from VS-IMSRG calculations based on EM1.8/2.0 interaction.


To investigate the origin of significant ISB with large MEDs and to further test the \textit{ab initio} calculations, we calculated the average occupations of valence nucleons above the inner core $^{16}$O. The results of the $4_1^+$ ground state and $1_{1,2}^+$ excited states in the proton-rich nucleus $^{22}$Al (neutron-rich nucleus $^{22}$F) are presented in  Fig.~\ref{fig5}(b). The proton-rich $^{22}$Al isotope is weakly bound, while the mirror nucleus $^{22}$F is deeply bound. All of the calculations reveal that the valence protons (neutrons) in the ground state of $^{22}$Al ($^{22}$F ) mainly occupy the $0d_{5/2}$ orbital.  As for the excited states,  the results with 1.8/2.0(EM), NNLO$\mathrm{_{opt}}$, and $NN+3N(\text{lnl})$ interactions show that the occupations of the $1s_{1/2}$ orbital in the $1_1^+$ mirror states are larger than those in the $1_2^+$ mirror states, implying that the occupied $s$ orbital, which is unbound in $^{22}$Al but deeply bound in $^{22}$F, is responsible for generating the TES, leading to a large MED value.
Thus, we can conclude that the occupations of $s$ wave dominantly drive the significant ISB with a large MED in mirror nuclei, in which the proton-rich nucleus is weakly bound or unbound but the neutron-rich nucleus is deeply bound. 
However, the calculations with NNLO$_{\text{sat}}$ interaction show that the $1^+_1$ states of $^{22}$Al/$^{22}$F bear a relatively smaller $1s_{1/2}$ average occupation than those of the $1^{+}_2$ states, and the resultant  MED of the $1_1^+$ states is smaller than that of the $1_2^+$ states, inconsistent with experimental data. Therefore, further optimizations and higher order interaction of chiral EFT are needed for the NNLO$_{\text{sat}}$ interaction to reproduce the experimental data of $^{22}$Al. 

Our calculations are consistent with Ref.~\cite{PhysRevLett.125.192503} concerning the mechanism of large shift in the $^{22}$Al-$^{22}$F mirror states. It is known that the halo structure is related to the occupation of low-$l$ single particle orbit by the valance nucleons~\cite{tanihata_recent_2013}, leading to a larger spatial extension in the matter distribution. However, a small $\pi$s$_{1/2}$ occupation is obtained for the ground-state of $^{22}$Al in our ab initio calculations. Therefore, it should be the $1_1^+$ excited state, rather than the ground-state of $^{22}$Al, that could have a proton-halo structure.


\section{Summary}
In conclusion, the ground-state mass of $\mathrm{^{22}Al}$ was measured for the first time with high accuracy using the B$\mathrm{\rho}$-defined IMS technique in CSRe. With the new mass data, the MEDs of the two low-lying $\mathrm{1^{+}}$ states in $^{22}$Al and $^{22}$F mirror nuclei were determined, and their uncertainties were significantly reduced from 403~keV to 51 keV. \textit{Ab initio} VS-IMSRG calculations using four sets of nuclear interactions, derived from chiral effective field theory,  were performed as benchmark tests on the odd-odd nuclei $^{22}$Al and $^{22}$F. The results show that the valence-space in-medium similarity renormalization group calculations with 1.8/2.0(EM), NNLO$\mathrm{_{opt}}$ and $NN + 3N (\text{lnl})$ interactions well describe the MEDs of $1^+_{1,2}$ states in $^{22}$Al-$^{22}$F mirror nuclei, however, the results from NNLO$\mathrm{_{sat}}$ interaction are inconsistent with experimental data. Further calculations of the average occupations of single-particle valance orbits reveal that the state bearing a large MED and exhibiting Thomas-Erhman shift is mainly driven by the occupation of $s$-wave, which is unbound in the proton-rich nucleus, while being deeply bound in the neutron-rich mirror partner. The significant occupation of $\pi s_{1/2}$ orbit supports the suggested halo structure in the 1$^{+}_{1}$ state of $^{22}$Al~\cite{PhysRevLett.125.192503} which needs further experimental investigations.

\textit{Acknowledgments.}~
We thank the staff of the accelerator division of IMP for providing the stable $^{36}$Ar beam. This work is supported in part by the Strategic Priority Research Program of Chinese Academy of Sciences (Grant No. XDB34000000), the CAS Project for Young Scientists in Basic Research (Grant No. YSBR-002),  the NSFC (Grants No. 12135017, No. 12121005, No. 11975280, No. 12105333, No. 12205340, No. 12322507, No. 12305126, No. 12305151), the Gansu Natural Science Foundation under Grant No. 22JR5RA123 and 23JRRA614, and the National Key R$\&$D Program of China (Grant No. 2021YFA1601500). Y. M. X. and X. X. acknowledge the support from the Youth Innovation Promotion Association of Chinese Academy of Sciences (Grants No. 2021419 and No. 2022423). X. L. Y. acknowledges the support from Young Scholar of Regional Development, CAS ([2023]15).  We acknowledge the Gansu Advanced Computing Center for providing computational resources.

\bibliography{22Al}

\begin{thebibliography}{59}%
\makeatletter
\providecommand \@ifxundefined [1]{%
 \@ifx{#1\undefined}
}%
\providecommand \@ifnum [1]{%
 \ifnum #1\expandafter \@firstoftwo
 \else \expandafter \@secondoftwo
 \fi
}%
\providecommand \@ifx [1]{%
 \ifx #1\expandafter \@firstoftwo
 \else \expandafter \@secondoftwo
 \fi
}%
\providecommand \natexlab [1]{#1}%
\providecommand \enquote  [1]{``#1''}%
\providecommand \bibnamefont  [1]{#1}%
\providecommand \bibfnamefont [1]{#1}%
\providecommand \citenamefont [1]{#1}%
\providecommand \href@noop [0]{\@secondoftwo}%
\providecommand \href [0]{\begingroup \@sanitize@url \@href}%
\providecommand \@href[1]{\@@startlink{#1}\@@href}%
\providecommand \@@href[1]{\endgroup#1\@@endlink}%
\providecommand \@sanitize@url [0]{\catcode `\\12\catcode `\$12\catcode
  `\&12\catcode `\#12\catcode `\^12\catcode `\_12\catcode `\%12\relax}%
\providecommand \@@startlink[1]{}%
\providecommand \@@endlink[0]{}%
\providecommand \url  [0]{\begingroup\@sanitize@url \@url }%
\providecommand \@url [1]{\endgroup\@href {#1}{\urlprefix }}%
\providecommand \urlprefix  [0]{URL }%
\providecommand \Eprint [0]{\href }%
\providecommand \doibase [0]{http://dx.doi.org/}%
\providecommand \selectlanguage [0]{\@gobble}%
\providecommand \bibinfo  [0]{\@secondoftwo}%
\providecommand \bibfield  [0]{\@secondoftwo}%
\providecommand \translation [1]{[#1]}%
\providecommand \BibitemOpen [0]{}%
\providecommand \bibitemStop [0]{}%
\providecommand \bibitemNoStop [0]{.\EOS\space}%
\providecommand \EOS [0]{\spacefactor3000\relax}%
\providecommand \BibitemShut  [1]{\csname bibitem#1\endcsname}%
\let\auto@bib@innerbib\@empty
\bibitem [{\citenamefont {Hergert}\ \emph {et~al.}(2016)\citenamefont
  {Hergert}, \citenamefont {Bogner}, \citenamefont {Morris}, \citenamefont
  {Schwenk},\ and\ \citenamefont {Tsukiyama}}]{HERGERT2016165}%
  \BibitemOpen
  \bibfield  {author} {\bibinfo {author} {\bibfnamefont {H.}~\bibnamefont
  {Hergert}}, \bibinfo {author} {\bibfnamefont {S.}~\bibnamefont {Bogner}},
  \bibinfo {author} {\bibfnamefont {T.}~\bibnamefont {Morris}}, \bibinfo
  {author} {\bibfnamefont {A.}~\bibnamefont {Schwenk}}, \ and\ \bibinfo
  {author} {\bibfnamefont {K.}~\bibnamefont {Tsukiyama}},\ }\href {\doibase
  https://doi.org/10.1016/j.physrep.2015.12.007} {\bibfield  {journal}
  {\bibinfo  {journal} {Phys. Rep.}\ }\textbf {\bibinfo {volume} {621}},\
  \bibinfo {pages} {165} (\bibinfo {year} {2016})},\ \bibinfo {note} {memorial
  Volume in Honor of Gerald E. Brown}\BibitemShut {NoStop}%
\bibitem [{\citenamefont {Stroberg}\ \emph {et~al.}(2019)\citenamefont
  {Stroberg}, \citenamefont {Hergert}, \citenamefont {Bogner},\ and\
  \citenamefont {Holt}}]{doi:10.1146/annurev-nucl-101917-021120}%
  \BibitemOpen
  \bibfield  {author} {\bibinfo {author} {\bibfnamefont {S.~R.}\ \bibnamefont
  {Stroberg}}, \bibinfo {author} {\bibfnamefont {H.}~\bibnamefont {Hergert}},
  \bibinfo {author} {\bibfnamefont {S.~K.}\ \bibnamefont {Bogner}}, \ and\
  \bibinfo {author} {\bibfnamefont {J.~D.}\ \bibnamefont {Holt}},\ }\href
  {\doibase 10.1146/annurev-nucl-101917-021120} {\bibfield  {journal} {\bibinfo
   {journal} {Ann. Rev. Nucl. Part. Sci}\ }\textbf {\bibinfo {volume} {69}},\
  \bibinfo {pages} {307} (\bibinfo {year} {2019})}\BibitemShut {NoStop}%
\bibitem [{\citenamefont {Garcia~Ruiz}\ \emph {et~al.}(2016)\citenamefont
  {Garcia~Ruiz}, \citenamefont {Bissell}, \citenamefont {Blaum}, \citenamefont
  {Ekström}, \citenamefont {Frömmgen}, \citenamefont {Hagen}, \citenamefont
  {Hammen}, \citenamefont {Hebeler}, \citenamefont {Holt}, \citenamefont
  {Jansen}, \citenamefont {Kowalska}, \citenamefont {Kreim}, \citenamefont
  {Nazarewicz}, \citenamefont {Neugart}, \citenamefont {Neyens}, \citenamefont
  {Nörtershäuser}, \citenamefont {Papenbrock}, \citenamefont {Papuga},
  \citenamefont {Schwenk}, \citenamefont {Simonis}, \citenamefont {Wendt},\
  and\ \citenamefont {Yordanov}}]{3}%
  \BibitemOpen
  \bibfield  {author} {\bibinfo {author} {\bibfnamefont {R.~F.}\ \bibnamefont
  {Garcia~Ruiz}}, \bibinfo {author} {\bibfnamefont {M.~L.}\ \bibnamefont
  {Bissell}}, \bibinfo {author} {\bibfnamefont {K.}~\bibnamefont {Blaum}},
  \bibinfo {author} {\bibfnamefont {A.}~\bibnamefont {Ekström}}, \bibinfo
  {author} {\bibfnamefont {N.}~\bibnamefont {Frömmgen}}, \bibinfo {author}
  {\bibfnamefont {G.}~\bibnamefont {Hagen}}, \bibinfo {author} {\bibfnamefont
  {M.}~\bibnamefont {Hammen}}, \bibinfo {author} {\bibfnamefont
  {K.}~\bibnamefont {Hebeler}}, \bibinfo {author} {\bibfnamefont {J.~D.}\
  \bibnamefont {Holt}}, \bibinfo {author} {\bibfnamefont {G.~R.}\ \bibnamefont
  {Jansen}}, \bibinfo {author} {\bibfnamefont {M.}~\bibnamefont {Kowalska}},
  \bibinfo {author} {\bibfnamefont {K.}~\bibnamefont {Kreim}}, \bibinfo
  {author} {\bibfnamefont {W.}~\bibnamefont {Nazarewicz}}, \bibinfo {author}
  {\bibfnamefont {R.}~\bibnamefont {Neugart}}, \bibinfo {author} {\bibfnamefont
  {G.}~\bibnamefont {Neyens}}, \bibinfo {author} {\bibfnamefont
  {W.}~\bibnamefont {Nörtershäuser}}, \bibinfo {author} {\bibfnamefont
  {T.}~\bibnamefont {Papenbrock}}, \bibinfo {author} {\bibfnamefont
  {J.}~\bibnamefont {Papuga}}, \bibinfo {author} {\bibfnamefont
  {A.}~\bibnamefont {Schwenk}}, \bibinfo {author} {\bibfnamefont
  {J.}~\bibnamefont {Simonis}}, \bibinfo {author} {\bibfnamefont {K.~A.}\
  \bibnamefont {Wendt}}, \ and\ \bibinfo {author} {\bibfnamefont {D.~T.}\
  \bibnamefont {Yordanov}},\ }\href {\doibase 10.1038/nphys3645} {\bibfield
  {journal} {\bibinfo  {journal} {Nat. Phys.}\ }\textbf {\bibinfo {volume}
  {12}},\ \bibinfo {pages} {594} (\bibinfo {year} {2016})}\BibitemShut
  {NoStop}%
\bibitem [{\citenamefont {Mougeot}\ \emph {et~al.}(2021)\citenamefont
  {Mougeot}, \citenamefont {Atanasov}, \citenamefont {Karthein}, \citenamefont
  {Wolf}, \citenamefont {Ascher}, \citenamefont {Blaum}, \citenamefont
  {Chrysalidis}, \citenamefont {Hagen}, \citenamefont {Holt}, \citenamefont
  {Huang}, \citenamefont {Jansen}, \citenamefont {Kulikov}, \citenamefont
  {Litvinov}, \citenamefont {Lunney}, \citenamefont {Manea}, \citenamefont
  {Miyagi}, \citenamefont {Papenbrock}, \citenamefont {Schweikhard},
  \citenamefont {Schwenk}, \citenamefont {Steinsberger}, \citenamefont
  {Stroberg}, \citenamefont {Sun}, \citenamefont {Welker}, \citenamefont
  {Wienholtz}, \citenamefont {Wilkins},\ and\ \citenamefont {Zuber}}]{5}%
  \BibitemOpen
  \bibfield  {author} {\bibinfo {author} {\bibfnamefont {M.}~\bibnamefont
  {Mougeot}}, \bibinfo {author} {\bibfnamefont {D.}~\bibnamefont {Atanasov}},
  \bibinfo {author} {\bibfnamefont {J.}~\bibnamefont {Karthein}}, \bibinfo
  {author} {\bibfnamefont {R.~N.}\ \bibnamefont {Wolf}}, \bibinfo {author}
  {\bibfnamefont {P.}~\bibnamefont {Ascher}}, \bibinfo {author} {\bibfnamefont
  {K.}~\bibnamefont {Blaum}}, \bibinfo {author} {\bibfnamefont
  {K.}~\bibnamefont {Chrysalidis}}, \bibinfo {author} {\bibfnamefont
  {G.}~\bibnamefont {Hagen}}, \bibinfo {author} {\bibfnamefont {J.~D.}\
  \bibnamefont {Holt}}, \bibinfo {author} {\bibfnamefont {W.~J.}\ \bibnamefont
  {Huang}}, \bibinfo {author} {\bibfnamefont {G.~R.}\ \bibnamefont {Jansen}},
  \bibinfo {author} {\bibfnamefont {I.}~\bibnamefont {Kulikov}}, \bibinfo
  {author} {\bibfnamefont {Y.~A.}\ \bibnamefont {Litvinov}}, \bibinfo {author}
  {\bibfnamefont {D.}~\bibnamefont {Lunney}}, \bibinfo {author} {\bibfnamefont
  {V.}~\bibnamefont {Manea}}, \bibinfo {author} {\bibfnamefont
  {T.}~\bibnamefont {Miyagi}}, \bibinfo {author} {\bibfnamefont
  {T.}~\bibnamefont {Papenbrock}}, \bibinfo {author} {\bibfnamefont
  {L.}~\bibnamefont {Schweikhard}}, \bibinfo {author} {\bibfnamefont
  {A.}~\bibnamefont {Schwenk}}, \bibinfo {author} {\bibfnamefont
  {T.}~\bibnamefont {Steinsberger}}, \bibinfo {author} {\bibfnamefont {S.~R.}\
  \bibnamefont {Stroberg}}, \bibinfo {author} {\bibfnamefont {Z.~H.}\
  \bibnamefont {Sun}}, \bibinfo {author} {\bibfnamefont {A.}~\bibnamefont
  {Welker}}, \bibinfo {author} {\bibfnamefont {F.}~\bibnamefont {Wienholtz}},
  \bibinfo {author} {\bibfnamefont {S.~G.}\ \bibnamefont {Wilkins}}, \ and\
  \bibinfo {author} {\bibfnamefont {K.}~\bibnamefont {Zuber}},\ }\href
  {\doibase 10.1038/s41567-021-01326-9} {\bibfield  {journal} {\bibinfo
  {journal} {Nat. Phys.}\ }\textbf {\bibinfo {volume} {17}},\ \bibinfo {pages}
  {1099} (\bibinfo {year} {2021})}\BibitemShut {NoStop}%
\bibitem [{\citenamefont {Stroberg}\ \emph {et~al.}(2017)\citenamefont
  {Stroberg}, \citenamefont {Calci}, \citenamefont {Hergert}, \citenamefont
  {Holt}, \citenamefont {Bogner}, \citenamefont {Roth},\ and\ \citenamefont
  {Schwenk}}]{4}%
  \BibitemOpen
  \bibfield  {author} {\bibinfo {author} {\bibfnamefont {S.~R.}\ \bibnamefont
  {Stroberg}}, \bibinfo {author} {\bibfnamefont {A.}~\bibnamefont {Calci}},
  \bibinfo {author} {\bibfnamefont {H.}~\bibnamefont {Hergert}}, \bibinfo
  {author} {\bibfnamefont {J.~D.}\ \bibnamefont {Holt}}, \bibinfo {author}
  {\bibfnamefont {S.~K.}\ \bibnamefont {Bogner}}, \bibinfo {author}
  {\bibfnamefont {R.}~\bibnamefont {Roth}}, \ and\ \bibinfo {author}
  {\bibfnamefont {A.}~\bibnamefont {Schwenk}},\ }\href {\doibase
  10.1103/PhysRevLett.118.032502} {\bibfield  {journal} {\bibinfo  {journal}
  {Phys. Rev. Lett.}\ }\textbf {\bibinfo {volume} {118}},\ \bibinfo {pages}
  {032502} (\bibinfo {year} {2017})}\BibitemShut {NoStop}%
\bibitem [{\citenamefont {Morris}\ \emph {et~al.}(2018)\citenamefont {Morris},
  \citenamefont {Simonis}, \citenamefont {Stroberg}, \citenamefont {Stumpf},
  \citenamefont {Hagen}, \citenamefont {Holt}, \citenamefont {Jansen},
  \citenamefont {Papenbrock}, \citenamefont {Roth},\ and\ \citenamefont
  {Schwenk}}]{31}%
  \BibitemOpen
  \bibfield  {author} {\bibinfo {author} {\bibfnamefont {T.~D.}\ \bibnamefont
  {Morris}}, \bibinfo {author} {\bibfnamefont {J.}~\bibnamefont {Simonis}},
  \bibinfo {author} {\bibfnamefont {S.~R.}\ \bibnamefont {Stroberg}}, \bibinfo
  {author} {\bibfnamefont {C.}~\bibnamefont {Stumpf}}, \bibinfo {author}
  {\bibfnamefont {G.}~\bibnamefont {Hagen}}, \bibinfo {author} {\bibfnamefont
  {J.~D.}\ \bibnamefont {Holt}}, \bibinfo {author} {\bibfnamefont {G.~R.}\
  \bibnamefont {Jansen}}, \bibinfo {author} {\bibfnamefont {T.}~\bibnamefont
  {Papenbrock}}, \bibinfo {author} {\bibfnamefont {R.}~\bibnamefont {Roth}}, \
  and\ \bibinfo {author} {\bibfnamefont {A.}~\bibnamefont {Schwenk}},\ }\href
  {\doibase 10.1103/PhysRevLett.120.152503} {\bibfield  {journal} {\bibinfo
  {journal} {Phys. Rev. Lett.}\ }\textbf {\bibinfo {volume} {120}},\ \bibinfo
  {pages} {152503} (\bibinfo {year} {2018})}\BibitemShut {NoStop}%
\bibitem [{\citenamefont {Hagen}\ \emph {et~al.}(2016)\citenamefont {Hagen},
  \citenamefont {Ekström}, \citenamefont {Forssén}, \citenamefont {Jansen},
  \citenamefont {Nazarewicz}, \citenamefont {Papenbrock}, \citenamefont
  {Wendt}, \citenamefont {Bacca}, \citenamefont {Barnea}, \citenamefont
  {Carlsson}, \citenamefont {Drischler}, \citenamefont {Hebeler}, \citenamefont
  {Hjorth-Jensen}, \citenamefont {Miorelli}, \citenamefont {Orlandini},
  \citenamefont {Schwenk},\ and\ \citenamefont {Simonis}}]{33}%
  \BibitemOpen
  \bibfield  {author} {\bibinfo {author} {\bibfnamefont {G.}~\bibnamefont
  {Hagen}}, \bibinfo {author} {\bibfnamefont {A.}~\bibnamefont {Ekström}},
  \bibinfo {author} {\bibfnamefont {C.}~\bibnamefont {Forssén}}, \bibinfo
  {author} {\bibfnamefont {G.~R.}\ \bibnamefont {Jansen}}, \bibinfo {author}
  {\bibfnamefont {W.}~\bibnamefont {Nazarewicz}}, \bibinfo {author}
  {\bibfnamefont {T.}~\bibnamefont {Papenbrock}}, \bibinfo {author}
  {\bibfnamefont {K.~A.}\ \bibnamefont {Wendt}}, \bibinfo {author}
  {\bibfnamefont {S.}~\bibnamefont {Bacca}}, \bibinfo {author} {\bibfnamefont
  {N.}~\bibnamefont {Barnea}}, \bibinfo {author} {\bibfnamefont
  {B.}~\bibnamefont {Carlsson}}, \bibinfo {author} {\bibfnamefont
  {C.}~\bibnamefont {Drischler}}, \bibinfo {author} {\bibfnamefont
  {K.}~\bibnamefont {Hebeler}}, \bibinfo {author} {\bibfnamefont
  {M.}~\bibnamefont {Hjorth-Jensen}}, \bibinfo {author} {\bibfnamefont
  {M.}~\bibnamefont {Miorelli}}, \bibinfo {author} {\bibfnamefont
  {G.}~\bibnamefont {Orlandini}}, \bibinfo {author} {\bibfnamefont
  {A.}~\bibnamefont {Schwenk}}, \ and\ \bibinfo {author} {\bibfnamefont
  {J.}~\bibnamefont {Simonis}},\ }\href {\doibase 10.1038/nphys3529} {\bibfield
   {journal} {\bibinfo  {journal} {Nat. Phys.}\ }\textbf {\bibinfo {volume}
  {12}},\ \bibinfo {pages} {186} (\bibinfo {year} {2016})}\BibitemShut
  {NoStop}%
\bibitem [{\citenamefont {Wang}\ \emph {et~al.}(2023)\citenamefont {Wang},
  \citenamefont {Zhang}, \citenamefont {Zhou}, \citenamefont {Zhou},
  \citenamefont {Xu}, \citenamefont {Liu}, \citenamefont {Li}, \citenamefont
  {Niu}, \citenamefont {Huang}, \citenamefont {Yuan}, \citenamefont {Zhang},
  \citenamefont {Xu}, \citenamefont {Litvinov}, \citenamefont {Blaum},
  \citenamefont {Meisel}, \citenamefont {Casten}, \citenamefont {Cakirli},
  \citenamefont {Chen}, \citenamefont {Deng}, \citenamefont {Fu}, \citenamefont
  {Ge}, \citenamefont {Li}, \citenamefont {Liao}, \citenamefont {Litvinov},
  \citenamefont {Shuai}, \citenamefont {Shi}, \citenamefont {Song},
  \citenamefont {Sun}, \citenamefont {Wang}, \citenamefont {Xing},
  \citenamefont {Xu}, \citenamefont {Yan}, \citenamefont {Yang}, \citenamefont
  {Yuan}, \citenamefont {Zeng},\ and\ \citenamefont {Zhang}}]{wang_mass_2023}%
  \BibitemOpen
  \bibfield  {author} {\bibinfo {author} {\bibfnamefont {M.}~\bibnamefont
  {Wang}}, \bibinfo {author} {\bibfnamefont {Y.~H.}\ \bibnamefont {Zhang}},
  \bibinfo {author} {\bibfnamefont {X.}~\bibnamefont {Zhou}}, \bibinfo {author}
  {\bibfnamefont {X.~H.}\ \bibnamefont {Zhou}}, \bibinfo {author}
  {\bibfnamefont {H.~S.}\ \bibnamefont {Xu}}, \bibinfo {author} {\bibfnamefont
  {M.~L.}\ \bibnamefont {Liu}}, \bibinfo {author} {\bibfnamefont {J.~G.}\
  \bibnamefont {Li}}, \bibinfo {author} {\bibfnamefont {Y.~F.}\ \bibnamefont
  {Niu}}, \bibinfo {author} {\bibfnamefont {W.~J.}\ \bibnamefont {Huang}},
  \bibinfo {author} {\bibfnamefont {Q.}~\bibnamefont {Yuan}}, \bibinfo {author}
  {\bibfnamefont {S.}~\bibnamefont {Zhang}}, \bibinfo {author} {\bibfnamefont
  {F.~R.}\ \bibnamefont {Xu}}, \bibinfo {author} {\bibfnamefont {Y.~A.}\
  \bibnamefont {Litvinov}}, \bibinfo {author} {\bibfnamefont {K.}~\bibnamefont
  {Blaum}}, \bibinfo {author} {\bibfnamefont {Z.}~\bibnamefont {Meisel}},
  \bibinfo {author} {\bibfnamefont {R.~F.}\ \bibnamefont {Casten}}, \bibinfo
  {author} {\bibfnamefont {R.~B.}\ \bibnamefont {Cakirli}}, \bibinfo {author}
  {\bibfnamefont {R.~J.}\ \bibnamefont {Chen}}, \bibinfo {author}
  {\bibfnamefont {H.~Y.}\ \bibnamefont {Deng}}, \bibinfo {author}
  {\bibfnamefont {C.~Y.}\ \bibnamefont {Fu}}, \bibinfo {author} {\bibfnamefont
  {W.~W.}\ \bibnamefont {Ge}}, \bibinfo {author} {\bibfnamefont {H.~F.}\
  \bibnamefont {Li}}, \bibinfo {author} {\bibfnamefont {T.}~\bibnamefont
  {Liao}}, \bibinfo {author} {\bibfnamefont {S.~A.}\ \bibnamefont {Litvinov}},
  \bibinfo {author} {\bibfnamefont {P.}~\bibnamefont {Shuai}}, \bibinfo
  {author} {\bibfnamefont {J.~Y.}\ \bibnamefont {Shi}}, \bibinfo {author}
  {\bibfnamefont {Y.~N.}\ \bibnamefont {Song}}, \bibinfo {author}
  {\bibfnamefont {M.~Z.}\ \bibnamefont {Sun}}, \bibinfo {author} {\bibfnamefont
  {Q.}~\bibnamefont {Wang}}, \bibinfo {author} {\bibfnamefont {Y.~M.}\
  \bibnamefont {Xing}}, \bibinfo {author} {\bibfnamefont {X.}~\bibnamefont
  {Xu}}, \bibinfo {author} {\bibfnamefont {X.~L.}\ \bibnamefont {Yan}},
  \bibinfo {author} {\bibfnamefont {J.~C.}\ \bibnamefont {Yang}}, \bibinfo
  {author} {\bibfnamefont {Y.~J.}\ \bibnamefont {Yuan}}, \bibinfo {author}
  {\bibfnamefont {Q.}~\bibnamefont {Zeng}}, \ and\ \bibinfo {author}
  {\bibfnamefont {M.}~\bibnamefont {Zhang}},\ }\href {\doibase
  10.1103/PhysRevLett.130.192501} {\bibfield  {journal} {\bibinfo  {journal}
  {Phys. Rev. Lett.}\ }\textbf {\bibinfo {volume} {130}},\ \bibinfo {pages}
  {192501} (\bibinfo {year} {2023})}\BibitemShut {NoStop}%
\bibitem [{\citenamefont {Stroberg}\ \emph {et~al.}(2021)\citenamefont
  {Stroberg}, \citenamefont {Holt}, \citenamefont {Schwenk},\ and\
  \citenamefont {Simonis}}]{PhysRevLett.126.022501}%
  \BibitemOpen
  \bibfield  {author} {\bibinfo {author} {\bibfnamefont {S.~R.}\ \bibnamefont
  {Stroberg}}, \bibinfo {author} {\bibfnamefont {J.~D.}\ \bibnamefont {Holt}},
  \bibinfo {author} {\bibfnamefont {A.}~\bibnamefont {Schwenk}}, \ and\
  \bibinfo {author} {\bibfnamefont {J.}~\bibnamefont {Simonis}},\ }\href
  {\doibase 10.1103/PhysRevLett.126.022501} {\bibfield  {journal} {\bibinfo
  {journal} {Phys. Rev. Lett.}\ }\textbf {\bibinfo {volume} {126}},\ \bibinfo
  {pages} {022501} (\bibinfo {year} {2021})}\BibitemShut {NoStop}%
\bibitem [{\citenamefont {Hebeler}(2021)}]{6}%
  \BibitemOpen
  \bibfield  {author} {\bibinfo {author} {\bibfnamefont {K.}~\bibnamefont
  {Hebeler}},\ }\href@noop {} {\bibfield  {journal} {\bibinfo  {journal} {Phys.
  Rep.}\ }\textbf {\bibinfo {volume} {890}},\ \bibinfo {pages} {1} (\bibinfo
  {year} {2021})}\BibitemShut {NoStop}%
\bibitem [{\citenamefont {Li}\ \emph {et~al.}(2023{\natexlab{a}})\citenamefont
  {Li}, \citenamefont {Yuan}, \citenamefont {Li}, \citenamefont {Xie},
  \citenamefont {Zhang}, \citenamefont {Zhang}, \citenamefont {Xu},
  \citenamefont {Michel}, \citenamefont {Xu},\ and\ \citenamefont
  {Zuo}}]{PhysRevC.107.014302}%
  \BibitemOpen
  \bibfield  {author} {\bibinfo {author} {\bibfnamefont {H.~H.}\ \bibnamefont
  {Li}}, \bibinfo {author} {\bibfnamefont {Q.}~\bibnamefont {Yuan}}, \bibinfo
  {author} {\bibfnamefont {J.~G.}\ \bibnamefont {Li}}, \bibinfo {author}
  {\bibfnamefont {M.~R.}\ \bibnamefont {Xie}}, \bibinfo {author} {\bibfnamefont
  {S.}~\bibnamefont {Zhang}}, \bibinfo {author} {\bibfnamefont {Y.~H.}\
  \bibnamefont {Zhang}}, \bibinfo {author} {\bibfnamefont {X.~X.}\ \bibnamefont
  {Xu}}, \bibinfo {author} {\bibfnamefont {N.}~\bibnamefont {Michel}}, \bibinfo
  {author} {\bibfnamefont {F.~R.}\ \bibnamefont {Xu}}, \ and\ \bibinfo {author}
  {\bibfnamefont {W.}~\bibnamefont {Zuo}},\ }\href {\doibase
  10.1103/PhysRevC.107.014302} {\bibfield  {journal} {\bibinfo  {journal}
  {Phys. Rev. C}\ }\textbf {\bibinfo {volume} {107}},\ \bibinfo {pages}
  {014302} (\bibinfo {year} {2023}{\natexlab{a}})}\BibitemShut {NoStop}%
\bibitem [{\citenamefont {Zhang}\ \emph {et~al.}()\citenamefont {Zhang},
  \citenamefont {Ma}, \citenamefont {Li}, \citenamefont {Hu}, \citenamefont
  {Yuan}, \citenamefont {Cheng},\ and\ \citenamefont {Xu}}]{zhang_roles_2022}%
  \BibitemOpen
  \bibfield  {author} {\bibinfo {author} {\bibfnamefont {S.}~\bibnamefont
  {Zhang}}, \bibinfo {author} {\bibfnamefont {Y.~Z.}\ \bibnamefont {Ma}},
  \bibinfo {author} {\bibfnamefont {J.~G.}\ \bibnamefont {Li}}, \bibinfo
  {author} {\bibfnamefont {B.~S.}\ \bibnamefont {Hu}}, \bibinfo {author}
  {\bibfnamefont {Q.}~\bibnamefont {Yuan}}, \bibinfo {author} {\bibfnamefont
  {Z.~H.}\ \bibnamefont {Cheng}}, \ and\ \bibinfo {author} {\bibfnamefont
  {F.~R.}\ \bibnamefont {Xu}},\ }\href {\doibase
  10.1016/j.physletb.2022.136958} {\ \textbf {\bibinfo {volume} {827}},\
  \bibinfo {pages} {136958}}\BibitemShut {NoStop}%
\bibitem [{\citenamefont {Li}\ \emph {et~al.}(2023{\natexlab{b}})\citenamefont
  {Li}, \citenamefont {Li}, \citenamefont {Zhang}, \citenamefont {Xing},\ and\
  \citenamefont {Zuo}}]{LI2023138197}%
  \BibitemOpen
  \bibfield  {author} {\bibinfo {author} {\bibfnamefont {J.~G.}\ \bibnamefont
  {Li}}, \bibinfo {author} {\bibfnamefont {H.~H.}\ \bibnamefont {Li}}, \bibinfo
  {author} {\bibfnamefont {S.}~\bibnamefont {Zhang}}, \bibinfo {author}
  {\bibfnamefont {Y.~M.}\ \bibnamefont {Xing}}, \ and\ \bibinfo {author}
  {\bibfnamefont {W.}~\bibnamefont {Zuo}},\ }\href {\doibase
  https://doi.org/10.1016/j.physletb.2023.138197} {\bibfield  {journal}
  {\bibinfo  {journal} {Phys. Lett. B}\ }\textbf {\bibinfo {volume} {846}},\
  \bibinfo {pages} {138197} (\bibinfo {year} {2023}{\natexlab{b}})}\BibitemShut
  {NoStop}%
\bibitem [{\citenamefont {Hu}\ \emph {et~al.}()\citenamefont {Hu},
  \citenamefont {Wu}, \citenamefont {Li}, \citenamefont {Ma}, \citenamefont
  {Sun}, \citenamefont {Michel},\ and\ \citenamefont {Xu}}]{hu_ab-initio_2020}%
  \BibitemOpen
  \bibfield  {author} {\bibinfo {author} {\bibfnamefont {B.~S.}\ \bibnamefont
  {Hu}}, \bibinfo {author} {\bibfnamefont {Q.}~\bibnamefont {Wu}}, \bibinfo
  {author} {\bibfnamefont {J.~G.}\ \bibnamefont {Li}}, \bibinfo {author}
  {\bibfnamefont {Y.~Z.}\ \bibnamefont {Ma}}, \bibinfo {author} {\bibfnamefont
  {Z.~H.}\ \bibnamefont {Sun}}, \bibinfo {author} {\bibfnamefont
  {N.}~\bibnamefont {Michel}}, \ and\ \bibinfo {author} {\bibfnamefont {F.~R.}\
  \bibnamefont {Xu}},\ }\href {\doibase
  https://doi.org/10.1016/j.physletb.2020.135206} {\bibfield  {journal}
  {\bibinfo  {journal} {Phys. Lett. B}\ }\textbf {\bibinfo {volume} {802}},\
  \bibinfo {pages} {135206}}\BibitemShut {NoStop}%
\bibitem [{\citenamefont {Lepailleur}\ \emph {et~al.}(2013)\citenamefont
  {Lepailleur}, \citenamefont {Sorlin}, \citenamefont {Caceres}, \citenamefont
  {Bastin}, \citenamefont {Borcea}, \citenamefont {Borcea}, \citenamefont
  {Brown}, \citenamefont {Gaudefroy}, \citenamefont {Gr\'evy}, \citenamefont
  {Grinyer}, \citenamefont {Hagen}, \citenamefont {Hjorth-Jensen},
  \citenamefont {Jansen}, \citenamefont {Llidoo}, \citenamefont {Negoita},
  \citenamefont {de~Oliveira}, \citenamefont {Porquet}, \citenamefont {Rotaru},
  \citenamefont {Saint-Laurent}, \citenamefont {Sohler}, \citenamefont
  {Stanoiu},\ and\ \citenamefont {Thomas}}]{PhysRevLett.110.082502}%
  \BibitemOpen
  \bibfield  {author} {\bibinfo {author} {\bibfnamefont {A.}~\bibnamefont
  {Lepailleur}}, \bibinfo {author} {\bibfnamefont {O.}~\bibnamefont {Sorlin}},
  \bibinfo {author} {\bibfnamefont {L.}~\bibnamefont {Caceres}}, \bibinfo
  {author} {\bibfnamefont {B.}~\bibnamefont {Bastin}}, \bibinfo {author}
  {\bibfnamefont {C.}~\bibnamefont {Borcea}}, \bibinfo {author} {\bibfnamefont
  {R.}~\bibnamefont {Borcea}}, \bibinfo {author} {\bibfnamefont {B.~A.}\
  \bibnamefont {Brown}}, \bibinfo {author} {\bibfnamefont {L.}~\bibnamefont
  {Gaudefroy}}, \bibinfo {author} {\bibfnamefont {S.}~\bibnamefont {Gr\'evy}},
  \bibinfo {author} {\bibfnamefont {G.~F.}\ \bibnamefont {Grinyer}}, \bibinfo
  {author} {\bibfnamefont {G.}~\bibnamefont {Hagen}}, \bibinfo {author}
  {\bibfnamefont {M.}~\bibnamefont {Hjorth-Jensen}}, \bibinfo {author}
  {\bibfnamefont {G.~R.}\ \bibnamefont {Jansen}}, \bibinfo {author}
  {\bibfnamefont {O.}~\bibnamefont {Llidoo}}, \bibinfo {author} {\bibfnamefont
  {F.}~\bibnamefont {Negoita}}, \bibinfo {author} {\bibfnamefont
  {F.}~\bibnamefont {de~Oliveira}}, \bibinfo {author} {\bibfnamefont {M.-G.}\
  \bibnamefont {Porquet}}, \bibinfo {author} {\bibfnamefont {F.}~\bibnamefont
  {Rotaru}}, \bibinfo {author} {\bibfnamefont {M.-G.}\ \bibnamefont
  {Saint-Laurent}}, \bibinfo {author} {\bibfnamefont {D.}~\bibnamefont
  {Sohler}}, \bibinfo {author} {\bibfnamefont {M.}~\bibnamefont {Stanoiu}}, \
  and\ \bibinfo {author} {\bibfnamefont {J.~C.}\ \bibnamefont {Thomas}},\
  }\href {\doibase 10.1103/PhysRevLett.110.082502} {\bibfield  {journal}
  {\bibinfo  {journal} {Phys. Rev. Lett.}\ }\textbf {\bibinfo {volume} {110}},\
  \bibinfo {pages} {082502} (\bibinfo {year} {2013})}\BibitemShut {NoStop}%
\bibitem [{\citenamefont {Vandebrouck}\ \emph {et~al.}(2017)\citenamefont
  {Vandebrouck}, \citenamefont {Lepailleur}, \citenamefont {Sorlin},
  \citenamefont {Aumann}, \citenamefont {Caesar}, \citenamefont {Holl},
  \citenamefont {Panin}, \citenamefont {Wamers}, \citenamefont {Stroberg},
  \citenamefont {Holt}, \citenamefont {de~Oliveira~Santos}, \citenamefont
  {Alvarez-Pol}, \citenamefont {Atar}, \citenamefont {Avdeichikov},
  \citenamefont {Beceiro-Novo}, \citenamefont {Bemmerer}, \citenamefont
  {Benlliure}, \citenamefont {Bertulani}, \citenamefont {Bogner}, \citenamefont
  {Boillos}, \citenamefont {Boretzky}, \citenamefont {Borge}, \citenamefont
  {Caama\~no}, \citenamefont {Casarejos}, \citenamefont {Catford},
  \citenamefont {Cederk\"all}, \citenamefont {Chartier}, \citenamefont
  {Chulkov}, \citenamefont {Cortina-Gil}, \citenamefont {Cravo}, \citenamefont
  {Crespo}, \citenamefont {Datta~Pramanik}, \citenamefont
  {D\'{\i}az~Fern\'andez}, \citenamefont {Dillmann}, \citenamefont {Elekes},
  \citenamefont {Enders}, \citenamefont {Ershova}, \citenamefont {Estrad\'e},
  \citenamefont {Farinon}, \citenamefont {Fraile}, \citenamefont {Freer},
  \citenamefont {Galaviz}, \citenamefont {Geissel}, \citenamefont
  {Gernh\"auser}, \citenamefont {Gibelin}, \citenamefont {Golubev},
  \citenamefont {G\"obel}, \citenamefont {Hagdahl}, \citenamefont {Heftrich},
  \citenamefont {Heil}, \citenamefont {Heine}, \citenamefont {Heinz},
  \citenamefont {Henriques}, \citenamefont {Hergert}, \citenamefont {Hufnagel},
  \citenamefont {Ignatov}, \citenamefont {Johansson}, \citenamefont {Jonson},
  \citenamefont {Kahlbow}, \citenamefont {Kalantar-Nayestanaki}, \citenamefont
  {Kanungo}, \citenamefont {Kelic-Heil}, \citenamefont {Knyazev}, \citenamefont
  {Kr\"oll}, \citenamefont {Kurz}, \citenamefont {Labiche}, \citenamefont
  {Langer}, \citenamefont {Le~Bleis}, \citenamefont {Lemmon}, \citenamefont
  {Lindberg}, \citenamefont {Machado}, \citenamefont {Marganiec}, \citenamefont
  {Marqu\'es}, \citenamefont {Movsesyan}, \citenamefont {Nacher}, \citenamefont
  {Najafi}, \citenamefont {Nikolskii}, \citenamefont {Nilsson}, \citenamefont
  {Nociforo}, \citenamefont {Paschalis}, \citenamefont {Perea}, \citenamefont
  {Petri}, \citenamefont {Pietri}, \citenamefont {Plag}, \citenamefont
  {Reifarth}, \citenamefont {Ribeiro}, \citenamefont {Rigollet}, \citenamefont
  {R\"oder}, \citenamefont {Rossi}, \citenamefont {Savran}, \citenamefont
  {Scheit}, \citenamefont {Schwenk}, \citenamefont {Simon}, \citenamefont
  {Syndikus}, \citenamefont {Taylor}, \citenamefont {Tengblad}, \citenamefont
  {Thies}, \citenamefont {Togano}, \citenamefont {Velho}, \citenamefont
  {Volkov}, \citenamefont {Wagner}, \citenamefont {Weick}, \citenamefont
  {Wheldon}, \citenamefont {Wilson}, \citenamefont {Winfield}, \citenamefont
  {Woods}, \citenamefont {Yakorev}, \citenamefont {Zhukov}, \citenamefont
  {Zilges},\ and\ \citenamefont {Zuber}}]{PhysRevC.96.054305}%
  \BibitemOpen
  \bibfield  {author} {\bibinfo {author} {\bibfnamefont {M.}~\bibnamefont
  {Vandebrouck}}, \bibinfo {author} {\bibfnamefont {A.}~\bibnamefont
  {Lepailleur}}, \bibinfo {author} {\bibfnamefont {O.}~\bibnamefont {Sorlin}},
  \bibinfo {author} {\bibfnamefont {T.}~\bibnamefont {Aumann}}, \bibinfo
  {author} {\bibfnamefont {C.}~\bibnamefont {Caesar}}, \bibinfo {author}
  {\bibfnamefont {M.}~\bibnamefont {Holl}}, \bibinfo {author} {\bibfnamefont
  {V.}~\bibnamefont {Panin}}, \bibinfo {author} {\bibfnamefont
  {F.}~\bibnamefont {Wamers}}, \bibinfo {author} {\bibfnamefont {S.~R.}\
  \bibnamefont {Stroberg}}, \bibinfo {author} {\bibfnamefont {J.~D.}\
  \bibnamefont {Holt}}, \bibinfo {author} {\bibfnamefont {F.}~\bibnamefont
  {de~Oliveira~Santos}}, \bibinfo {author} {\bibfnamefont {H.}~\bibnamefont
  {Alvarez-Pol}}, \bibinfo {author} {\bibfnamefont {L.}~\bibnamefont {Atar}},
  \bibinfo {author} {\bibfnamefont {V.}~\bibnamefont {Avdeichikov}}, \bibinfo
  {author} {\bibfnamefont {S.}~\bibnamefont {Beceiro-Novo}}, \bibinfo {author}
  {\bibfnamefont {D.}~\bibnamefont {Bemmerer}}, \bibinfo {author}
  {\bibfnamefont {J.}~\bibnamefont {Benlliure}}, \bibinfo {author}
  {\bibfnamefont {C.~A.}\ \bibnamefont {Bertulani}}, \bibinfo {author}
  {\bibfnamefont {S.~K.}\ \bibnamefont {Bogner}}, \bibinfo {author}
  {\bibfnamefont {J.~M.}\ \bibnamefont {Boillos}}, \bibinfo {author}
  {\bibfnamefont {K.}~\bibnamefont {Boretzky}}, \bibinfo {author}
  {\bibfnamefont {M.~J.~G.}\ \bibnamefont {Borge}}, \bibinfo {author}
  {\bibfnamefont {M.}~\bibnamefont {Caama\~no}}, \bibinfo {author}
  {\bibfnamefont {E.}~\bibnamefont {Casarejos}}, \bibinfo {author}
  {\bibfnamefont {W.}~\bibnamefont {Catford}}, \bibinfo {author} {\bibfnamefont
  {J.}~\bibnamefont {Cederk\"all}}, \bibinfo {author} {\bibfnamefont
  {M.}~\bibnamefont {Chartier}}, \bibinfo {author} {\bibfnamefont
  {L.}~\bibnamefont {Chulkov}}, \bibinfo {author} {\bibfnamefont
  {D.}~\bibnamefont {Cortina-Gil}}, \bibinfo {author} {\bibfnamefont
  {E.}~\bibnamefont {Cravo}}, \bibinfo {author} {\bibfnamefont
  {R.}~\bibnamefont {Crespo}}, \bibinfo {author} {\bibfnamefont
  {U.}~\bibnamefont {Datta~Pramanik}}, \bibinfo {author} {\bibfnamefont
  {P.}~\bibnamefont {D\'{\i}az~Fern\'andez}}, \bibinfo {author} {\bibfnamefont
  {I.}~\bibnamefont {Dillmann}}, \bibinfo {author} {\bibfnamefont
  {Z.}~\bibnamefont {Elekes}}, \bibinfo {author} {\bibfnamefont
  {J.}~\bibnamefont {Enders}}, \bibinfo {author} {\bibfnamefont
  {O.}~\bibnamefont {Ershova}}, \bibinfo {author} {\bibfnamefont
  {A.}~\bibnamefont {Estrad\'e}}, \bibinfo {author} {\bibfnamefont
  {F.}~\bibnamefont {Farinon}}, \bibinfo {author} {\bibfnamefont {L.~M.}\
  \bibnamefont {Fraile}}, \bibinfo {author} {\bibfnamefont {M.}~\bibnamefont
  {Freer}}, \bibinfo {author} {\bibfnamefont {D.}~\bibnamefont {Galaviz}},
  \bibinfo {author} {\bibfnamefont {H.}~\bibnamefont {Geissel}}, \bibinfo
  {author} {\bibfnamefont {R.}~\bibnamefont {Gernh\"auser}}, \bibinfo {author}
  {\bibfnamefont {J.}~\bibnamefont {Gibelin}}, \bibinfo {author} {\bibfnamefont
  {P.}~\bibnamefont {Golubev}}, \bibinfo {author} {\bibfnamefont
  {K.}~\bibnamefont {G\"obel}}, \bibinfo {author} {\bibfnamefont
  {J.}~\bibnamefont {Hagdahl}}, \bibinfo {author} {\bibfnamefont
  {T.}~\bibnamefont {Heftrich}}, \bibinfo {author} {\bibfnamefont
  {M.}~\bibnamefont {Heil}}, \bibinfo {author} {\bibfnamefont {M.}~\bibnamefont
  {Heine}}, \bibinfo {author} {\bibfnamefont {A.}~\bibnamefont {Heinz}},
  \bibinfo {author} {\bibfnamefont {A.}~\bibnamefont {Henriques}}, \bibinfo
  {author} {\bibfnamefont {H.}~\bibnamefont {Hergert}}, \bibinfo {author}
  {\bibfnamefont {A.}~\bibnamefont {Hufnagel}}, \bibinfo {author}
  {\bibfnamefont {A.}~\bibnamefont {Ignatov}}, \bibinfo {author} {\bibfnamefont
  {H.~T.}\ \bibnamefont {Johansson}}, \bibinfo {author} {\bibfnamefont
  {B.}~\bibnamefont {Jonson}}, \bibinfo {author} {\bibfnamefont
  {J.}~\bibnamefont {Kahlbow}}, \bibinfo {author} {\bibfnamefont
  {N.}~\bibnamefont {Kalantar-Nayestanaki}}, \bibinfo {author} {\bibfnamefont
  {R.}~\bibnamefont {Kanungo}}, \bibinfo {author} {\bibfnamefont
  {A.}~\bibnamefont {Kelic-Heil}}, \bibinfo {author} {\bibfnamefont
  {A.}~\bibnamefont {Knyazev}}, \bibinfo {author} {\bibfnamefont
  {T.}~\bibnamefont {Kr\"oll}}, \bibinfo {author} {\bibfnamefont
  {N.}~\bibnamefont {Kurz}}, \bibinfo {author} {\bibfnamefont {M.}~\bibnamefont
  {Labiche}}, \bibinfo {author} {\bibfnamefont {C.}~\bibnamefont {Langer}},
  \bibinfo {author} {\bibfnamefont {T.}~\bibnamefont {Le~Bleis}}, \bibinfo
  {author} {\bibfnamefont {R.}~\bibnamefont {Lemmon}}, \bibinfo {author}
  {\bibfnamefont {S.}~\bibnamefont {Lindberg}}, \bibinfo {author}
  {\bibfnamefont {J.}~\bibnamefont {Machado}}, \bibinfo {author} {\bibfnamefont
  {J.}~\bibnamefont {Marganiec}}, \bibinfo {author} {\bibfnamefont {F.~M.}\
  \bibnamefont {Marqu\'es}}, \bibinfo {author} {\bibfnamefont {A.}~\bibnamefont
  {Movsesyan}}, \bibinfo {author} {\bibfnamefont {E.}~\bibnamefont {Nacher}},
  \bibinfo {author} {\bibfnamefont {M.}~\bibnamefont {Najafi}}, \bibinfo
  {author} {\bibfnamefont {E.}~\bibnamefont {Nikolskii}}, \bibinfo {author}
  {\bibfnamefont {T.}~\bibnamefont {Nilsson}}, \bibinfo {author} {\bibfnamefont
  {C.}~\bibnamefont {Nociforo}}, \bibinfo {author} {\bibfnamefont
  {S.}~\bibnamefont {Paschalis}}, \bibinfo {author} {\bibfnamefont
  {A.}~\bibnamefont {Perea}}, \bibinfo {author} {\bibfnamefont
  {M.}~\bibnamefont {Petri}}, \bibinfo {author} {\bibfnamefont
  {S.}~\bibnamefont {Pietri}}, \bibinfo {author} {\bibfnamefont
  {R.}~\bibnamefont {Plag}}, \bibinfo {author} {\bibfnamefont {R.}~\bibnamefont
  {Reifarth}}, \bibinfo {author} {\bibfnamefont {G.}~\bibnamefont {Ribeiro}},
  \bibinfo {author} {\bibfnamefont {C.}~\bibnamefont {Rigollet}}, \bibinfo
  {author} {\bibfnamefont {M.}~\bibnamefont {R\"oder}}, \bibinfo {author}
  {\bibfnamefont {D.}~\bibnamefont {Rossi}}, \bibinfo {author} {\bibfnamefont
  {D.}~\bibnamefont {Savran}}, \bibinfo {author} {\bibfnamefont
  {H.}~\bibnamefont {Scheit}}, \bibinfo {author} {\bibfnamefont
  {A.}~\bibnamefont {Schwenk}}, \bibinfo {author} {\bibfnamefont
  {H.}~\bibnamefont {Simon}}, \bibinfo {author} {\bibfnamefont
  {I.}~\bibnamefont {Syndikus}}, \bibinfo {author} {\bibfnamefont {J.~T.}\
  \bibnamefont {Taylor}}, \bibinfo {author} {\bibfnamefont {O.}~\bibnamefont
  {Tengblad}}, \bibinfo {author} {\bibfnamefont {R.}~\bibnamefont {Thies}},
  \bibinfo {author} {\bibfnamefont {Y.}~\bibnamefont {Togano}}, \bibinfo
  {author} {\bibfnamefont {P.}~\bibnamefont {Velho}}, \bibinfo {author}
  {\bibfnamefont {V.}~\bibnamefont {Volkov}}, \bibinfo {author} {\bibfnamefont
  {A.}~\bibnamefont {Wagner}}, \bibinfo {author} {\bibfnamefont
  {H.}~\bibnamefont {Weick}}, \bibinfo {author} {\bibfnamefont
  {C.}~\bibnamefont {Wheldon}}, \bibinfo {author} {\bibfnamefont
  {G.}~\bibnamefont {Wilson}}, \bibinfo {author} {\bibfnamefont {J.~S.}\
  \bibnamefont {Winfield}}, \bibinfo {author} {\bibfnamefont {P.}~\bibnamefont
  {Woods}}, \bibinfo {author} {\bibfnamefont {D.}~\bibnamefont {Yakorev}},
  \bibinfo {author} {\bibfnamefont {M.}~\bibnamefont {Zhukov}}, \bibinfo
  {author} {\bibfnamefont {A.}~\bibnamefont {Zilges}}, \ and\ \bibinfo {author}
  {\bibfnamefont {K.}~\bibnamefont {Zuber}} (\bibinfo {collaboration} {R3B
  collaboration}),\ }\href {\doibase 10.1103/PhysRevC.96.054305} {\bibfield
  {journal} {\bibinfo  {journal} {Phys. Rev. C}\ }\textbf {\bibinfo {volume}
  {96}},\ \bibinfo {pages} {054305} (\bibinfo {year} {2017})}\BibitemShut
  {NoStop}%
\bibitem [{\citenamefont {Zuker}\ \emph {et~al.}(2002)\citenamefont {Zuker},
  \citenamefont {Lenzi}, \citenamefont {Mart\'{\i}nez-Pinedo},\ and\
  \citenamefont {Poves}}]{PhysRevLett.89.142502}%
  \BibitemOpen
  \bibfield  {author} {\bibinfo {author} {\bibfnamefont {A.~P.}\ \bibnamefont
  {Zuker}}, \bibinfo {author} {\bibfnamefont {S.~M.}\ \bibnamefont {Lenzi}},
  \bibinfo {author} {\bibfnamefont {G.}~\bibnamefont {Mart\'{\i}nez-Pinedo}}, \
  and\ \bibinfo {author} {\bibfnamefont {A.}~\bibnamefont {Poves}},\ }\href
  {\doibase 10.1103/PhysRevLett.89.142502} {\bibfield  {journal} {\bibinfo
  {journal} {Phys. Rev. Lett.}\ }\textbf {\bibinfo {volume} {89}},\ \bibinfo
  {pages} {142502} (\bibinfo {year} {2002})}\BibitemShut {NoStop}%
\bibitem [{\citenamefont {Ekman}\ \emph {et~al.}(2004)\citenamefont {Ekman},
  \citenamefont {Rudolph}, \citenamefont {Fahlander}, \citenamefont {Zuker},
  \citenamefont {Bentley}, \citenamefont {Lenzi}, \citenamefont {Andreoiu},
  \citenamefont {Axiotis}, \citenamefont {de~Angelis}, \citenamefont {Farnea},
  \citenamefont {Gadea}, \citenamefont {Kr\"oll}, \citenamefont
  {M\ifmmode~\u{a}\else \u{a}\fi{}rginean}, \citenamefont {Martinez},
  \citenamefont {Mineva}, \citenamefont {Rossi-Alvarez},\ and\ \citenamefont
  {Ur}}]{PhysRevLett.92.132502}%
  \BibitemOpen
  \bibfield  {author} {\bibinfo {author} {\bibfnamefont {J.}~\bibnamefont
  {Ekman}}, \bibinfo {author} {\bibfnamefont {D.}~\bibnamefont {Rudolph}},
  \bibinfo {author} {\bibfnamefont {C.}~\bibnamefont {Fahlander}}, \bibinfo
  {author} {\bibfnamefont {A.~P.}\ \bibnamefont {Zuker}}, \bibinfo {author}
  {\bibfnamefont {M.~A.}\ \bibnamefont {Bentley}}, \bibinfo {author}
  {\bibfnamefont {S.~M.}\ \bibnamefont {Lenzi}}, \bibinfo {author}
  {\bibfnamefont {C.}~\bibnamefont {Andreoiu}}, \bibinfo {author}
  {\bibfnamefont {M.}~\bibnamefont {Axiotis}}, \bibinfo {author} {\bibfnamefont
  {G.}~\bibnamefont {de~Angelis}}, \bibinfo {author} {\bibfnamefont
  {E.}~\bibnamefont {Farnea}}, \bibinfo {author} {\bibfnamefont
  {A.}~\bibnamefont {Gadea}}, \bibinfo {author} {\bibfnamefont
  {T.}~\bibnamefont {Kr\"oll}}, \bibinfo {author} {\bibfnamefont
  {N.}~\bibnamefont {M\ifmmode~\u{a}\else \u{a}\fi{}rginean}}, \bibinfo
  {author} {\bibfnamefont {T.}~\bibnamefont {Martinez}}, \bibinfo {author}
  {\bibfnamefont {M.~N.}\ \bibnamefont {Mineva}}, \bibinfo {author}
  {\bibfnamefont {C.}~\bibnamefont {Rossi-Alvarez}}, \ and\ \bibinfo {author}
  {\bibfnamefont {C.~A.}\ \bibnamefont {Ur}},\ }\href {\doibase
  10.1103/PhysRevLett.92.132502} {\bibfield  {journal} {\bibinfo  {journal}
  {Phys. Rev. Lett.}\ }\textbf {\bibinfo {volume} {92}},\ \bibinfo {pages}
  {132502} (\bibinfo {year} {2004})}\BibitemShut {NoStop}%
\bibitem [{\citenamefont {Bentley}\ \emph {et~al.}(2006)\citenamefont
  {Bentley}, \citenamefont {Chandler}, \citenamefont {Taylor}, \citenamefont
  {Brown}, \citenamefont {Carpenter}, \citenamefont {Davids}, \citenamefont
  {Ekman}, \citenamefont {Freeman}, \citenamefont {Garrett}, \citenamefont
  {Hammond}, \citenamefont {Janssens}, \citenamefont {Lenzi}, \citenamefont
  {Lister}, \citenamefont {du~Rietz},\ and\ \citenamefont
  {Seweryniak}}]{PhysRevLett.97.132501}%
  \BibitemOpen
  \bibfield  {author} {\bibinfo {author} {\bibfnamefont {M.~A.}\ \bibnamefont
  {Bentley}}, \bibinfo {author} {\bibfnamefont {C.}~\bibnamefont {Chandler}},
  \bibinfo {author} {\bibfnamefont {M.~J.}\ \bibnamefont {Taylor}}, \bibinfo
  {author} {\bibfnamefont {J.~R.}\ \bibnamefont {Brown}}, \bibinfo {author}
  {\bibfnamefont {M.~P.}\ \bibnamefont {Carpenter}}, \bibinfo {author}
  {\bibfnamefont {C.}~\bibnamefont {Davids}}, \bibinfo {author} {\bibfnamefont
  {J.}~\bibnamefont {Ekman}}, \bibinfo {author} {\bibfnamefont {S.~J.}\
  \bibnamefont {Freeman}}, \bibinfo {author} {\bibfnamefont {P.~E.}\
  \bibnamefont {Garrett}}, \bibinfo {author} {\bibfnamefont {G.}~\bibnamefont
  {Hammond}}, \bibinfo {author} {\bibfnamefont {R.~V.~F.}\ \bibnamefont
  {Janssens}}, \bibinfo {author} {\bibfnamefont {S.~M.}\ \bibnamefont {Lenzi}},
  \bibinfo {author} {\bibfnamefont {C.~J.}\ \bibnamefont {Lister}}, \bibinfo
  {author} {\bibfnamefont {R.}~\bibnamefont {du~Rietz}}, \ and\ \bibinfo
  {author} {\bibfnamefont {D.}~\bibnamefont {Seweryniak}},\ }\href {\doibase
  10.1103/PhysRevLett.97.132501} {\bibfield  {journal} {\bibinfo  {journal}
  {Phys. Rev. Lett.}\ }\textbf {\bibinfo {volume} {97}},\ \bibinfo {pages}
  {132501} (\bibinfo {year} {2006})}\BibitemShut {NoStop}%
\bibitem [{\citenamefont {Gadea}\ \emph {et~al.}(2006)\citenamefont {Gadea},
  \citenamefont {Lenzi}, \citenamefont {Lunardi}, \citenamefont
  {M\ifmmode~\u{a}\else \u{a}\fi{}rginean}, \citenamefont {Zuker},
  \citenamefont {de~Angelis}, \citenamefont {Axiotis}, \citenamefont
  {Mart\'{\i}nez}, \citenamefont {Napoli}, \citenamefont {Farnea},
  \citenamefont {Menegazzo}, \citenamefont {Pavan}, \citenamefont {Ur},
  \citenamefont {Bazzacco}, \citenamefont {Venturelli}, \citenamefont
  {Kleinheinz}, \citenamefont {Bednarczyk}, \citenamefont {Curien},
  \citenamefont {Dorvaux}, \citenamefont {Nyberg}, \citenamefont {Grawe},
  \citenamefont {G\'orska}, \citenamefont {Palacz}, \citenamefont {Lagergren},
  \citenamefont {Milechina}, \citenamefont {Ekman}, \citenamefont {Rudolph},
  \citenamefont {Andreoiu}, \citenamefont {Bentley}, \citenamefont {Gelletly},
  \citenamefont {Rubio}, \citenamefont {Algora}, \citenamefont {Nacher},
  \citenamefont {Caballero}, \citenamefont {Trotta},\ and\ \citenamefont
  {Moszy\ifmmode~\acute{n}\else \'{n}\fi{}ski}}]{PhysRevLett.97.152501}%
  \BibitemOpen
  \bibfield  {author} {\bibinfo {author} {\bibfnamefont {A.}~\bibnamefont
  {Gadea}}, \bibinfo {author} {\bibfnamefont {S.~M.}\ \bibnamefont {Lenzi}},
  \bibinfo {author} {\bibfnamefont {S.}~\bibnamefont {Lunardi}}, \bibinfo
  {author} {\bibfnamefont {N.}~\bibnamefont {M\ifmmode~\u{a}\else
  \u{a}\fi{}rginean}}, \bibinfo {author} {\bibfnamefont {A.~P.}\ \bibnamefont
  {Zuker}}, \bibinfo {author} {\bibfnamefont {G.}~\bibnamefont {de~Angelis}},
  \bibinfo {author} {\bibfnamefont {M.}~\bibnamefont {Axiotis}}, \bibinfo
  {author} {\bibfnamefont {T.}~\bibnamefont {Mart\'{\i}nez}}, \bibinfo {author}
  {\bibfnamefont {D.~R.}\ \bibnamefont {Napoli}}, \bibinfo {author}
  {\bibfnamefont {E.}~\bibnamefont {Farnea}}, \bibinfo {author} {\bibfnamefont
  {R.}~\bibnamefont {Menegazzo}}, \bibinfo {author} {\bibfnamefont
  {P.}~\bibnamefont {Pavan}}, \bibinfo {author} {\bibfnamefont {C.~A.}\
  \bibnamefont {Ur}}, \bibinfo {author} {\bibfnamefont {D.}~\bibnamefont
  {Bazzacco}}, \bibinfo {author} {\bibfnamefont {R.}~\bibnamefont
  {Venturelli}}, \bibinfo {author} {\bibfnamefont {P.}~\bibnamefont
  {Kleinheinz}}, \bibinfo {author} {\bibfnamefont {P.}~\bibnamefont
  {Bednarczyk}}, \bibinfo {author} {\bibfnamefont {D.}~\bibnamefont {Curien}},
  \bibinfo {author} {\bibfnamefont {O.}~\bibnamefont {Dorvaux}}, \bibinfo
  {author} {\bibfnamefont {J.}~\bibnamefont {Nyberg}}, \bibinfo {author}
  {\bibfnamefont {H.}~\bibnamefont {Grawe}}, \bibinfo {author} {\bibfnamefont
  {M.}~\bibnamefont {G\'orska}}, \bibinfo {author} {\bibfnamefont
  {M.}~\bibnamefont {Palacz}}, \bibinfo {author} {\bibfnamefont
  {K.}~\bibnamefont {Lagergren}}, \bibinfo {author} {\bibfnamefont
  {L.}~\bibnamefont {Milechina}}, \bibinfo {author} {\bibfnamefont
  {J.}~\bibnamefont {Ekman}}, \bibinfo {author} {\bibfnamefont
  {D.}~\bibnamefont {Rudolph}}, \bibinfo {author} {\bibfnamefont
  {C.}~\bibnamefont {Andreoiu}}, \bibinfo {author} {\bibfnamefont {M.~A.}\
  \bibnamefont {Bentley}}, \bibinfo {author} {\bibfnamefont {W.}~\bibnamefont
  {Gelletly}}, \bibinfo {author} {\bibfnamefont {B.}~\bibnamefont {Rubio}},
  \bibinfo {author} {\bibfnamefont {A.}~\bibnamefont {Algora}}, \bibinfo
  {author} {\bibfnamefont {E.}~\bibnamefont {Nacher}}, \bibinfo {author}
  {\bibfnamefont {L.}~\bibnamefont {Caballero}}, \bibinfo {author}
  {\bibfnamefont {M.}~\bibnamefont {Trotta}}, \ and\ \bibinfo {author}
  {\bibfnamefont {M.}~\bibnamefont {Moszy\ifmmode~\acute{n}\else
  \'{n}\fi{}ski}},\ }\href {\doibase 10.1103/PhysRevLett.97.152501} {\bibfield
  {journal} {\bibinfo  {journal} {Phys. Rev. Lett.}\ }\textbf {\bibinfo
  {volume} {97}},\ \bibinfo {pages} {152501} (\bibinfo {year}
  {2006})}\BibitemShut {NoStop}%
\bibitem [{\citenamefont {Kaneko}\ \emph {et~al.}(2013)\citenamefont {Kaneko},
  \citenamefont {Sun}, \citenamefont {Mizusaki},\ and\ \citenamefont
  {Tazaki}}]{PhysRevLett.110.172505}%
  \BibitemOpen
  \bibfield  {author} {\bibinfo {author} {\bibfnamefont {K.}~\bibnamefont
  {Kaneko}}, \bibinfo {author} {\bibfnamefont {Y.}~\bibnamefont {Sun}},
  \bibinfo {author} {\bibfnamefont {T.}~\bibnamefont {Mizusaki}}, \ and\
  \bibinfo {author} {\bibfnamefont {S.}~\bibnamefont {Tazaki}},\ }\href
  {\doibase 10.1103/PhysRevLett.110.172505} {\bibfield  {journal} {\bibinfo
  {journal} {Phys. Rev. Lett.}\ }\textbf {\bibinfo {volume} {110}},\ \bibinfo
  {pages} {172505} (\bibinfo {year} {2013})}\BibitemShut {NoStop}%
\bibitem [{\citenamefont {Lee}\ \emph {et~al.}(2020)\citenamefont {Lee},
  \citenamefont {Xu}, \citenamefont {Kaneko}, \citenamefont {Sun},
  \citenamefont {Lin}, \citenamefont {Sun}, \citenamefont {Liang},
  \citenamefont {Li}, \citenamefont {Li}, \citenamefont {Wu}, \citenamefont
  {Fang}, \citenamefont {Wang}, \citenamefont {Yang}, \citenamefont {Yuan},
  \citenamefont {Lam}, \citenamefont {Wang}, \citenamefont {Wang},
  \citenamefont {Wang}, \citenamefont {Ma}, \citenamefont {Liu}, \citenamefont
  {Li}, \citenamefont {Zhao}, \citenamefont {Yang}, \citenamefont {Ma},
  \citenamefont {Wang}, \citenamefont {Zhong}, \citenamefont {Zhong},
  \citenamefont {Yang}, \citenamefont {Jia}, \citenamefont {Wen}, \citenamefont
  {Pan}, \citenamefont {Zang}, \citenamefont {Wang}, \citenamefont {Wu},
  \citenamefont {Luo}, \citenamefont {Wang}, \citenamefont {Li}, \citenamefont
  {Shi}, \citenamefont {Nie}, \citenamefont {Li}, \citenamefont {Li},
  \citenamefont {Ma}, \citenamefont {Hu}, \citenamefont {Shi}, \citenamefont
  {Jin}, \citenamefont {Huang}, \citenamefont {Bai}, \citenamefont {Zhou},
  \citenamefont {Ma}, \citenamefont {Duan}, \citenamefont {Jin}, \citenamefont
  {Gao}, \citenamefont {Zhou}, \citenamefont {Hu}, \citenamefont {Wang},
  \citenamefont {Liu}, \citenamefont {Chen},\ and\ \citenamefont
  {Ma}}]{PhysRevLett.125.192503}%
  \BibitemOpen
  \bibfield  {author} {\bibinfo {author} {\bibfnamefont {J.}~\bibnamefont
  {Lee}}, \bibinfo {author} {\bibfnamefont {X.~X.}\ \bibnamefont {Xu}},
  \bibinfo {author} {\bibfnamefont {K.}~\bibnamefont {Kaneko}}, \bibinfo
  {author} {\bibfnamefont {Y.}~\bibnamefont {Sun}}, \bibinfo {author}
  {\bibfnamefont {C.~J.}\ \bibnamefont {Lin}}, \bibinfo {author} {\bibfnamefont
  {L.~J.}\ \bibnamefont {Sun}}, \bibinfo {author} {\bibfnamefont {P.~F.}\
  \bibnamefont {Liang}}, \bibinfo {author} {\bibfnamefont {Z.~H.}\ \bibnamefont
  {Li}}, \bibinfo {author} {\bibfnamefont {J.}~\bibnamefont {Li}}, \bibinfo
  {author} {\bibfnamefont {H.~Y.}\ \bibnamefont {Wu}}, \bibinfo {author}
  {\bibfnamefont {D.~Q.}\ \bibnamefont {Fang}}, \bibinfo {author}
  {\bibfnamefont {J.~S.}\ \bibnamefont {Wang}}, \bibinfo {author}
  {\bibfnamefont {Y.~Y.}\ \bibnamefont {Yang}}, \bibinfo {author}
  {\bibfnamefont {C.~X.}\ \bibnamefont {Yuan}}, \bibinfo {author}
  {\bibfnamefont {Y.~H.}\ \bibnamefont {Lam}}, \bibinfo {author} {\bibfnamefont
  {Y.~T.}\ \bibnamefont {Wang}}, \bibinfo {author} {\bibfnamefont
  {K.}~\bibnamefont {Wang}}, \bibinfo {author} {\bibfnamefont {J.~G.}\
  \bibnamefont {Wang}}, \bibinfo {author} {\bibfnamefont {J.~B.}\ \bibnamefont
  {Ma}}, \bibinfo {author} {\bibfnamefont {J.~J.}\ \bibnamefont {Liu}},
  \bibinfo {author} {\bibfnamefont {P.~J.}\ \bibnamefont {Li}}, \bibinfo
  {author} {\bibfnamefont {Q.~Q.}\ \bibnamefont {Zhao}}, \bibinfo {author}
  {\bibfnamefont {L.}~\bibnamefont {Yang}}, \bibinfo {author} {\bibfnamefont
  {N.~R.}\ \bibnamefont {Ma}}, \bibinfo {author} {\bibfnamefont {D.~X.}\
  \bibnamefont {Wang}}, \bibinfo {author} {\bibfnamefont {F.~P.}\ \bibnamefont
  {Zhong}}, \bibinfo {author} {\bibfnamefont {S.~H.}\ \bibnamefont {Zhong}},
  \bibinfo {author} {\bibfnamefont {F.}~\bibnamefont {Yang}}, \bibinfo {author}
  {\bibfnamefont {H.~M.}\ \bibnamefont {Jia}}, \bibinfo {author} {\bibfnamefont
  {P.~W.}\ \bibnamefont {Wen}}, \bibinfo {author} {\bibfnamefont
  {M.}~\bibnamefont {Pan}}, \bibinfo {author} {\bibfnamefont {H.~L.}\
  \bibnamefont {Zang}}, \bibinfo {author} {\bibfnamefont {X.}~\bibnamefont
  {Wang}}, \bibinfo {author} {\bibfnamefont {C.~G.}\ \bibnamefont {Wu}},
  \bibinfo {author} {\bibfnamefont {D.~W.}\ \bibnamefont {Luo}}, \bibinfo
  {author} {\bibfnamefont {H.~W.}\ \bibnamefont {Wang}}, \bibinfo {author}
  {\bibfnamefont {C.}~\bibnamefont {Li}}, \bibinfo {author} {\bibfnamefont
  {C.~Z.}\ \bibnamefont {Shi}}, \bibinfo {author} {\bibfnamefont {M.~W.}\
  \bibnamefont {Nie}}, \bibinfo {author} {\bibfnamefont {X.~F.}\ \bibnamefont
  {Li}}, \bibinfo {author} {\bibfnamefont {H.}~\bibnamefont {Li}}, \bibinfo
  {author} {\bibfnamefont {P.}~\bibnamefont {Ma}}, \bibinfo {author}
  {\bibfnamefont {Q.}~\bibnamefont {Hu}}, \bibinfo {author} {\bibfnamefont
  {G.~Z.}\ \bibnamefont {Shi}}, \bibinfo {author} {\bibfnamefont {S.~L.}\
  \bibnamefont {Jin}}, \bibinfo {author} {\bibfnamefont {M.~R.}\ \bibnamefont
  {Huang}}, \bibinfo {author} {\bibfnamefont {Z.}~\bibnamefont {Bai}}, \bibinfo
  {author} {\bibfnamefont {Y.~J.}\ \bibnamefont {Zhou}}, \bibinfo {author}
  {\bibfnamefont {W.~H.}\ \bibnamefont {Ma}}, \bibinfo {author} {\bibfnamefont
  {F.~F.}\ \bibnamefont {Duan}}, \bibinfo {author} {\bibfnamefont {S.~Y.}\
  \bibnamefont {Jin}}, \bibinfo {author} {\bibfnamefont {Q.~R.}\ \bibnamefont
  {Gao}}, \bibinfo {author} {\bibfnamefont {X.~H.}\ \bibnamefont {Zhou}},
  \bibinfo {author} {\bibfnamefont {Z.~G.}\ \bibnamefont {Hu}}, \bibinfo
  {author} {\bibfnamefont {M.}~\bibnamefont {Wang}}, \bibinfo {author}
  {\bibfnamefont {M.~L.}\ \bibnamefont {Liu}}, \bibinfo {author} {\bibfnamefont
  {R.~F.}\ \bibnamefont {Chen}}, \ and\ \bibinfo {author} {\bibfnamefont
  {X.~W.}\ \bibnamefont {Ma}} (\bibinfo {collaboration} {RIBLL
  Collaboration}),\ }\href {\doibase 10.1103/PhysRevLett.125.192503} {\bibfield
   {journal} {\bibinfo  {journal} {Phys. Rev. Lett.}\ }\textbf {\bibinfo
  {volume} {125}},\ \bibinfo {pages} {192503} (\bibinfo {year}
  {2020})}\BibitemShut {NoStop}%
\bibitem [{\citenamefont {Li}\ \emph {et~al.}(2023{\natexlab{c}})\citenamefont
  {Li}, \citenamefont {Li}, \citenamefont {Xie},\ and\ \citenamefont
  {W.}}]{CPC:10.1088/1674-1137/acf035}%
  \BibitemOpen
  \bibfield  {author} {\bibinfo {author} {\bibfnamefont {H.~H.}\ \bibnamefont
  {Li}}, \bibinfo {author} {\bibfnamefont {J.~G.}\ \bibnamefont {Li}}, \bibinfo
  {author} {\bibfnamefont {M.~R.}\ \bibnamefont {Xie}}, \ and\ \bibinfo
  {author} {\bibfnamefont {Z.}~\bibnamefont {W.}},\ }\href {\doibase
  10.1088/1674-1137/acf035} {\bibfield  {journal} {\bibinfo  {journal} {Chin.
  Phys. C}\ }\textbf {\bibinfo {volume} {47}},\ \bibinfo {pages} {124101}
  (\bibinfo {year} {2023}{\natexlab{c}})}\BibitemShut {NoStop}%
\bibitem [{\citenamefont {{Miller}}\ \emph {et~al.}(2006)\citenamefont
  {{Miller}}, \citenamefont {{Opper}},\ and\ \citenamefont
  {{Stephenson}}}]{2006ARNPS..56..253M}%
  \BibitemOpen
  \bibfield  {author} {\bibinfo {author} {\bibfnamefont {G.~A.}\ \bibnamefont
  {{Miller}}}, \bibinfo {author} {\bibfnamefont {A.~K.}\ \bibnamefont
  {{Opper}}}, \ and\ \bibinfo {author} {\bibfnamefont {E.~J.}\ \bibnamefont
  {{Stephenson}}},\ }\href {\doibase 10.1146/annurev.nucl.56.080805.140446}
  {\bibfield  {journal} {\bibinfo  {journal} {Ann. Rev. Nucl. Part. Sci}\
  }\textbf {\bibinfo {volume} {56}},\ \bibinfo {pages} {253} (\bibinfo {year}
  {2006})}\BibitemShut {NoStop}%
\bibitem [{\citenamefont {Thomas}(1952)}]{PhysRev.88.1109}%
  \BibitemOpen
  \bibfield  {author} {\bibinfo {author} {\bibfnamefont {R.~G.}\ \bibnamefont
  {Thomas}},\ }\href {\doibase 10.1103/PhysRev.88.1109} {\bibfield  {journal}
  {\bibinfo  {journal} {Phys. Rev.}\ }\textbf {\bibinfo {volume} {88}},\
  \bibinfo {pages} {1109} (\bibinfo {year} {1952})}\BibitemShut {NoStop}%
\bibitem [{\citenamefont {Ehrman}(1951)}]{PhysRev.81.412}%
  \BibitemOpen
  \bibfield  {author} {\bibinfo {author} {\bibfnamefont {J.~B.}\ \bibnamefont
  {Ehrman}},\ }\href {\doibase 10.1103/PhysRev.81.412} {\bibfield  {journal}
  {\bibinfo  {journal} {Phys. Rev.}\ }\textbf {\bibinfo {volume} {81}},\
  \bibinfo {pages} {412} (\bibinfo {year} {1951})}\BibitemShut {NoStop}%
\bibitem [{\citenamefont {Wang}\ \emph {et~al.}(2021)\citenamefont {Wang},
  \citenamefont {Huang}, \citenamefont {Kondev}, \citenamefont {Audi},\ and\
  \citenamefont {Naimi}}]{18}%
  \BibitemOpen
  \bibfield  {author} {\bibinfo {author} {\bibfnamefont {M.}~\bibnamefont
  {Wang}}, \bibinfo {author} {\bibfnamefont {W.~J.}\ \bibnamefont {Huang}},
  \bibinfo {author} {\bibfnamefont {F.~G.}\ \bibnamefont {Kondev}}, \bibinfo
  {author} {\bibfnamefont {G.}~\bibnamefont {Audi}}, \ and\ \bibinfo {author}
  {\bibfnamefont {S.}~\bibnamefont {Naimi}},\ }\href {\doibase
  10.1088/1674-1137/abddaf} {\bibfield  {journal} {\bibinfo  {journal} {Chin.
  Phys. C}\ }\textbf {\bibinfo {volume} {45}},\ \bibinfo {pages} {030003}
  (\bibinfo {year} {2021})}\BibitemShut {NoStop}%
\bibitem [{end()}]{endsf}%
  \BibitemOpen
  \href@noop {} {}\bibinfo {howpublished}
  {\url{https://www.nndc.bnl.gov/ensdf/}}\BibitemShut {NoStop}%
\bibitem [{\citenamefont {Xu}\ \emph {et~al.}(2023)\citenamefont {Xu},
  \citenamefont {Zhang}, \citenamefont {Li}, \citenamefont {Jin}, \citenamefont
  {Yuan}, \citenamefont {Cheng}, \citenamefont {Michel},\ and\ \citenamefont
  {Xu}}]{PhysRevC.108.L031301}%
  \BibitemOpen
  \bibfield  {author} {\bibinfo {author} {\bibfnamefont {Z.~C.}\ \bibnamefont
  {Xu}}, \bibinfo {author} {\bibfnamefont {S.}~\bibnamefont {Zhang}}, \bibinfo
  {author} {\bibfnamefont {J.~G.}\ \bibnamefont {Li}}, \bibinfo {author}
  {\bibfnamefont {S.~L.}\ \bibnamefont {Jin}}, \bibinfo {author} {\bibfnamefont
  {Q.}~\bibnamefont {Yuan}}, \bibinfo {author} {\bibfnamefont {Z.~H.}\
  \bibnamefont {Cheng}}, \bibinfo {author} {\bibfnamefont {N.}~\bibnamefont
  {Michel}}, \ and\ \bibinfo {author} {\bibfnamefont {F.~R.}\ \bibnamefont
  {Xu}},\ }\href {\doibase 10.1103/PhysRevC.108.L031301} {\bibfield  {journal}
  {\bibinfo  {journal} {Phys. Rev. C}\ }\textbf {\bibinfo {volume} {108}},\
  \bibinfo {pages} {L031301} (\bibinfo {year} {2023})}\BibitemShut {NoStop}%
\bibitem [{\citenamefont {Wang}\ \emph {et~al.}(2022)\citenamefont {Wang},
  \citenamefont {Zhang}, \citenamefont {Zhou}, \citenamefont {Zhang},
  \citenamefont {Litvinov}, \citenamefont {Xu}, \citenamefont {Chen},
  \citenamefont {Deng}, \citenamefont {Fu}, \citenamefont {Ge}, \citenamefont
  {Li}, \citenamefont {Liao}, \citenamefont {Litvinov}, \citenamefont {Shuai},
  \citenamefont {Shi}, \citenamefont {Si}, \citenamefont {Sidhu}, \citenamefont
  {Song}, \citenamefont {Sun}, \citenamefont {Suzuki}, \citenamefont {Wang},
  \citenamefont {Xing}, \citenamefont {Xu}, \citenamefont {Yamaguchi},
  \citenamefont {Yan}, \citenamefont {Yang}, \citenamefont {Yuan},
  \citenamefont {Zeng},\ and\ \citenamefont {Zhou}}]{17}%
  \BibitemOpen
  \bibfield  {author} {\bibinfo {author} {\bibfnamefont {M.}~\bibnamefont
  {Wang}}, \bibinfo {author} {\bibfnamefont {M.}~\bibnamefont {Zhang}},
  \bibinfo {author} {\bibfnamefont {X.}~\bibnamefont {Zhou}}, \bibinfo {author}
  {\bibfnamefont {Y.~H.}\ \bibnamefont {Zhang}}, \bibinfo {author}
  {\bibfnamefont {Y.~A.}\ \bibnamefont {Litvinov}}, \bibinfo {author}
  {\bibfnamefont {H.~S.}\ \bibnamefont {Xu}}, \bibinfo {author} {\bibfnamefont
  {R.~J.}\ \bibnamefont {Chen}}, \bibinfo {author} {\bibfnamefont {H.~Y.}\
  \bibnamefont {Deng}}, \bibinfo {author} {\bibfnamefont {C.~Y.}\ \bibnamefont
  {Fu}}, \bibinfo {author} {\bibfnamefont {W.~W.}\ \bibnamefont {Ge}}, \bibinfo
  {author} {\bibfnamefont {H.~F.}\ \bibnamefont {Li}}, \bibinfo {author}
  {\bibfnamefont {T.}~\bibnamefont {Liao}}, \bibinfo {author} {\bibfnamefont
  {S.~A.}\ \bibnamefont {Litvinov}}, \bibinfo {author} {\bibfnamefont
  {P.}~\bibnamefont {Shuai}}, \bibinfo {author} {\bibfnamefont {J.~Y.}\
  \bibnamefont {Shi}}, \bibinfo {author} {\bibfnamefont {M.}~\bibnamefont
  {Si}}, \bibinfo {author} {\bibfnamefont {R.~S.}\ \bibnamefont {Sidhu}},
  \bibinfo {author} {\bibfnamefont {Y.~N.}\ \bibnamefont {Song}}, \bibinfo
  {author} {\bibfnamefont {M.~Z.}\ \bibnamefont {Sun}}, \bibinfo {author}
  {\bibfnamefont {S.}~\bibnamefont {Suzuki}}, \bibinfo {author} {\bibfnamefont
  {Q.}~\bibnamefont {Wang}}, \bibinfo {author} {\bibfnamefont {Y.~M.}\
  \bibnamefont {Xing}}, \bibinfo {author} {\bibfnamefont {X.}~\bibnamefont
  {Xu}}, \bibinfo {author} {\bibfnamefont {T.}~\bibnamefont {Yamaguchi}},
  \bibinfo {author} {\bibfnamefont {X.~L.}\ \bibnamefont {Yan}}, \bibinfo
  {author} {\bibfnamefont {J.~C.}\ \bibnamefont {Yang}}, \bibinfo {author}
  {\bibfnamefont {Y.~J.}\ \bibnamefont {Yuan}}, \bibinfo {author}
  {\bibfnamefont {Q.}~\bibnamefont {Zeng}}, \ and\ \bibinfo {author}
  {\bibfnamefont {X.~H.}\ \bibnamefont {Zhou}},\ }\href {\doibase
  10.1103/PhysRevC.106.L051301} {\bibfield  {journal} {\bibinfo  {journal}
  {Phys. Rev. C}\ }\textbf {\bibinfo {volume} {106}},\ \bibinfo {pages}
  {L051301} (\bibinfo {year} {2022})}\BibitemShut {NoStop}%
\bibitem [{\citenamefont {Zhang}\ \emph {et~al.}(2023)\citenamefont {Zhang},
  \citenamefont {Zhou}, \citenamefont {Wang}, \citenamefont {Zhang},
  \citenamefont {Litvinov}, \citenamefont {Xu}, \citenamefont {Chen},
  \citenamefont {Deng}, \citenamefont {Fu}, \citenamefont {Ge}, \citenamefont
  {Li}, \citenamefont {Liao}, \citenamefont {Litvinov}, \citenamefont {Shuai},
  \citenamefont {Shi}, \citenamefont {Sidhu}, \citenamefont {Song},
  \citenamefont {Sun}, \citenamefont {Suzuki}, \citenamefont {Wang},
  \citenamefont {Xing}, \citenamefont {Xu}, \citenamefont {Yamaguchi},
  \citenamefont {Yan}, \citenamefont {Yang}, \citenamefont {Yuan},
  \citenamefont {Zeng},\ and\ \citenamefont {Zhou}}]{zhang_brho_2023}%
  \BibitemOpen
  \bibfield  {author} {\bibinfo {author} {\bibfnamefont {M.}~\bibnamefont
  {Zhang}}, \bibinfo {author} {\bibfnamefont {X.}~\bibnamefont {Zhou}},
  \bibinfo {author} {\bibfnamefont {M.}~\bibnamefont {Wang}}, \bibinfo {author}
  {\bibfnamefont {Y.~H.}\ \bibnamefont {Zhang}}, \bibinfo {author}
  {\bibfnamefont {Y.~A.}\ \bibnamefont {Litvinov}}, \bibinfo {author}
  {\bibfnamefont {H.~S.}\ \bibnamefont {Xu}}, \bibinfo {author} {\bibfnamefont
  {R.~J.}\ \bibnamefont {Chen}}, \bibinfo {author} {\bibfnamefont {H.~Y.}\
  \bibnamefont {Deng}}, \bibinfo {author} {\bibfnamefont {C.~Y.}\ \bibnamefont
  {Fu}}, \bibinfo {author} {\bibfnamefont {W.~W.}\ \bibnamefont {Ge}}, \bibinfo
  {author} {\bibfnamefont {H.~F.}\ \bibnamefont {Li}}, \bibinfo {author}
  {\bibfnamefont {T.}~\bibnamefont {Liao}}, \bibinfo {author} {\bibfnamefont
  {S.~A.}\ \bibnamefont {Litvinov}}, \bibinfo {author} {\bibfnamefont
  {P.}~\bibnamefont {Shuai}}, \bibinfo {author} {\bibfnamefont {J.~Y.}\
  \bibnamefont {Shi}}, \bibinfo {author} {\bibfnamefont {R.~S.}\ \bibnamefont
  {Sidhu}}, \bibinfo {author} {\bibfnamefont {Y.~N.}\ \bibnamefont {Song}},
  \bibinfo {author} {\bibfnamefont {M.~Z.}\ \bibnamefont {Sun}}, \bibinfo
  {author} {\bibfnamefont {S.}~\bibnamefont {Suzuki}}, \bibinfo {author}
  {\bibfnamefont {Q.}~\bibnamefont {Wang}}, \bibinfo {author} {\bibfnamefont
  {Y.~M.}\ \bibnamefont {Xing}}, \bibinfo {author} {\bibfnamefont
  {X.}~\bibnamefont {Xu}}, \bibinfo {author} {\bibfnamefont {T.}~\bibnamefont
  {Yamaguchi}}, \bibinfo {author} {\bibfnamefont {X.~L.}\ \bibnamefont {Yan}},
  \bibinfo {author} {\bibfnamefont {J.~C.}\ \bibnamefont {Yang}}, \bibinfo
  {author} {\bibfnamefont {Y.~J.}\ \bibnamefont {Yuan}}, \bibinfo {author}
  {\bibfnamefont {Q.}~\bibnamefont {Zeng}}, \ and\ \bibinfo {author}
  {\bibfnamefont {X.~H.}\ \bibnamefont {Zhou}},\ }\href {\doibase
  10.1140/epja/s10050-023-00928-6} {\bibfield  {journal} {\bibinfo  {journal}
  {Eur. Phys. J. A}\ }\textbf {\bibinfo {volume} {59}},\ \bibinfo {pages} {27}
  (\bibinfo {year} {2023})}\BibitemShut {NoStop}%
\bibitem [{\citenamefont {Xia}\ \emph {et~al.}(2002)\citenamefont {Xia},
  \citenamefont {Zhan}, \citenamefont {Wei}, \citenamefont {Yuan},
  \citenamefont {Song}, \citenamefont {Zhang}, \citenamefont {Yang},
  \citenamefont {Yuan}, \citenamefont {Gao}, \citenamefont {Zhao},
  \citenamefont {Yang}, \citenamefont {Xiao}, \citenamefont {Man},
  \citenamefont {Dang}, \citenamefont {Cai}, \citenamefont {Wang},
  \citenamefont {Tang}, \citenamefont {Qiao}, \citenamefont {Rao},
  \citenamefont {He}, \citenamefont {Mao},\ and\ \citenamefont {Zhou}}]{9}%
  \BibitemOpen
  \bibfield  {author} {\bibinfo {author} {\bibfnamefont {J.~W.}\ \bibnamefont
  {Xia}}, \bibinfo {author} {\bibfnamefont {W.~L.}\ \bibnamefont {Zhan}},
  \bibinfo {author} {\bibfnamefont {B.~W.}\ \bibnamefont {Wei}}, \bibinfo
  {author} {\bibfnamefont {Y.~J.}\ \bibnamefont {Yuan}}, \bibinfo {author}
  {\bibfnamefont {M.~T.}\ \bibnamefont {Song}}, \bibinfo {author}
  {\bibfnamefont {W.~Z.}\ \bibnamefont {Zhang}}, \bibinfo {author}
  {\bibfnamefont {X.~D.}\ \bibnamefont {Yang}}, \bibinfo {author}
  {\bibfnamefont {P.}~\bibnamefont {Yuan}}, \bibinfo {author} {\bibfnamefont
  {D.~Q.}\ \bibnamefont {Gao}}, \bibinfo {author} {\bibfnamefont {H.~W.}\
  \bibnamefont {Zhao}}, \bibinfo {author} {\bibfnamefont {X.~T.}\ \bibnamefont
  {Yang}}, \bibinfo {author} {\bibfnamefont {G.~Q.}\ \bibnamefont {Xiao}},
  \bibinfo {author} {\bibfnamefont {K.~T.}\ \bibnamefont {Man}}, \bibinfo
  {author} {\bibfnamefont {J.~R.}\ \bibnamefont {Dang}}, \bibinfo {author}
  {\bibfnamefont {X.~H.}\ \bibnamefont {Cai}}, \bibinfo {author} {\bibfnamefont
  {Y.~F.}\ \bibnamefont {Wang}}, \bibinfo {author} {\bibfnamefont {J.~Y.}\
  \bibnamefont {Tang}}, \bibinfo {author} {\bibfnamefont {W.~M.}\ \bibnamefont
  {Qiao}}, \bibinfo {author} {\bibfnamefont {Y.~N.}\ \bibnamefont {Rao}},
  \bibinfo {author} {\bibfnamefont {Y.}~\bibnamefont {He}}, \bibinfo {author}
  {\bibfnamefont {L.~Z.}\ \bibnamefont {Mao}}, \ and\ \bibinfo {author}
  {\bibfnamefont {Z.~Z.}\ \bibnamefont {Zhou}},\ }\href {\doibase
  10.1016/S0168-9002(02)00475-8} {\bibfield  {journal} {\bibinfo  {journal}
  {Nucl. Instrum. Methods Phys. Res., Sect. A}\ }\textbf {\bibinfo {volume}
  {488}},\ \bibinfo {pages} {11} (\bibinfo {year} {2002})}\BibitemShut
  {NoStop}%
\bibitem [{\citenamefont {Zhan}\ \emph {et~al.}(2010)\citenamefont {Zhan},
  \citenamefont {Xu}, \citenamefont {Xiao}, \citenamefont {Xia}, \citenamefont
  {Zhao},\ and\ \citenamefont {Yuan}}]{10}%
  \BibitemOpen
  \bibfield  {author} {\bibinfo {author} {\bibfnamefont {W.~L.}\ \bibnamefont
  {Zhan}}, \bibinfo {author} {\bibfnamefont {H.~S.}\ \bibnamefont {Xu}},
  \bibinfo {author} {\bibfnamefont {G.~Q.}\ \bibnamefont {Xiao}}, \bibinfo
  {author} {\bibfnamefont {J.~W.}\ \bibnamefont {Xia}}, \bibinfo {author}
  {\bibfnamefont {H.~W.}\ \bibnamefont {Zhao}}, \ and\ \bibinfo {author}
  {\bibfnamefont {Y.~J.}\ \bibnamefont {Yuan}},\ }\href {\doibase
  10.1016/j.nuclphysa.2010.01.126} {\bibfield  {journal} {\bibinfo  {journal}
  {Nucl. Phys. A}\ }\bibinfo {series} {The 10th {International} {Conference} on
  {Nucleus}-{Nucleus} {Collisions} ({NN2009})},\ \textbf {\bibinfo {volume}
  {834}},\ \bibinfo {pages} {694c} (\bibinfo {year} {2010})}\BibitemShut
  {NoStop}%
\bibitem [{\citenamefont {Tu}\ \emph {et~al.}(2011{\natexlab{a}})\citenamefont
  {Tu}, \citenamefont {Xu}, \citenamefont {Wang}, \citenamefont {Zhang},
  \citenamefont {Litvinov}, \citenamefont {Sun}, \citenamefont {Schatz},
  \citenamefont {Zhou}, \citenamefont {Yuan}, \citenamefont {Xia},
  \citenamefont {Audi}, \citenamefont {Blaum}, \citenamefont {Du},
  \citenamefont {Geng}, \citenamefont {Hu}, \citenamefont {Huang},
  \citenamefont {Jin}, \citenamefont {Liu}, \citenamefont {Liu}, \citenamefont
  {Ma}, \citenamefont {Mao}, \citenamefont {Mei}, \citenamefont {Shuai},
  \citenamefont {Sun}, \citenamefont {Suzuki}, \citenamefont {Tang},
  \citenamefont {Wang}, \citenamefont {Wang}, \citenamefont {Xiao},
  \citenamefont {Xu}, \citenamefont {Yamaguchi}, \citenamefont {Yamaguchi},
  \citenamefont {Yan}, \citenamefont {Yang}, \citenamefont {Ye}, \citenamefont
  {Zang}, \citenamefont {Zhao}, \citenamefont {Zhao}, \citenamefont {Zhang},\
  and\ \citenamefont {Zhan}}]{tu_direct_2011}%
  \BibitemOpen
  \bibfield  {author} {\bibinfo {author} {\bibfnamefont {X.~L.}\ \bibnamefont
  {Tu}}, \bibinfo {author} {\bibfnamefont {H.~S.}\ \bibnamefont {Xu}}, \bibinfo
  {author} {\bibfnamefont {M.}~\bibnamefont {Wang}}, \bibinfo {author}
  {\bibfnamefont {Y.~H.}\ \bibnamefont {Zhang}}, \bibinfo {author}
  {\bibfnamefont {Y.~A.}\ \bibnamefont {Litvinov}}, \bibinfo {author}
  {\bibfnamefont {Y.}~\bibnamefont {Sun}}, \bibinfo {author} {\bibfnamefont
  {H.}~\bibnamefont {Schatz}}, \bibinfo {author} {\bibfnamefont {X.~H.}\
  \bibnamefont {Zhou}}, \bibinfo {author} {\bibfnamefont {Y.~J.}\ \bibnamefont
  {Yuan}}, \bibinfo {author} {\bibfnamefont {J.~W.}\ \bibnamefont {Xia}},
  \bibinfo {author} {\bibfnamefont {G.}~\bibnamefont {Audi}}, \bibinfo {author}
  {\bibfnamefont {K.}~\bibnamefont {Blaum}}, \bibinfo {author} {\bibfnamefont
  {C.~M.}\ \bibnamefont {Du}}, \bibinfo {author} {\bibfnamefont
  {P.}~\bibnamefont {Geng}}, \bibinfo {author} {\bibfnamefont {Z.~G.}\
  \bibnamefont {Hu}}, \bibinfo {author} {\bibfnamefont {W.~X.}\ \bibnamefont
  {Huang}}, \bibinfo {author} {\bibfnamefont {S.~L.}\ \bibnamefont {Jin}},
  \bibinfo {author} {\bibfnamefont {L.~X.}\ \bibnamefont {Liu}}, \bibinfo
  {author} {\bibfnamefont {Y.}~\bibnamefont {Liu}}, \bibinfo {author}
  {\bibfnamefont {X.}~\bibnamefont {Ma}}, \bibinfo {author} {\bibfnamefont
  {R.~S.}\ \bibnamefont {Mao}}, \bibinfo {author} {\bibfnamefont
  {B.}~\bibnamefont {Mei}}, \bibinfo {author} {\bibfnamefont {P.}~\bibnamefont
  {Shuai}}, \bibinfo {author} {\bibfnamefont {Z.~Y.}\ \bibnamefont {Sun}},
  \bibinfo {author} {\bibfnamefont {H.}~\bibnamefont {Suzuki}}, \bibinfo
  {author} {\bibfnamefont {S.~W.}\ \bibnamefont {Tang}}, \bibinfo {author}
  {\bibfnamefont {J.~S.}\ \bibnamefont {Wang}}, \bibinfo {author}
  {\bibfnamefont {S.~T.}\ \bibnamefont {Wang}}, \bibinfo {author}
  {\bibfnamefont {G.~Q.}\ \bibnamefont {Xiao}}, \bibinfo {author}
  {\bibfnamefont {X.}~\bibnamefont {Xu}}, \bibinfo {author} {\bibfnamefont
  {T.}~\bibnamefont {Yamaguchi}}, \bibinfo {author} {\bibfnamefont
  {Y.}~\bibnamefont {Yamaguchi}}, \bibinfo {author} {\bibfnamefont {X.~L.}\
  \bibnamefont {Yan}}, \bibinfo {author} {\bibfnamefont {J.~C.}\ \bibnamefont
  {Yang}}, \bibinfo {author} {\bibfnamefont {R.~P.}\ \bibnamefont {Ye}},
  \bibinfo {author} {\bibfnamefont {Y.~D.}\ \bibnamefont {Zang}}, \bibinfo
  {author} {\bibfnamefont {H.~W.}\ \bibnamefont {Zhao}}, \bibinfo {author}
  {\bibfnamefont {T.~C.}\ \bibnamefont {Zhao}}, \bibinfo {author}
  {\bibfnamefont {X.~Y.}\ \bibnamefont {Zhang}}, \ and\ \bibinfo {author}
  {\bibfnamefont {W.~L.}\ \bibnamefont {Zhan}},\ }\href {\doibase
  10.1103/PhysRevLett.106.112501} {\bibfield  {journal} {\bibinfo  {journal}
  {Phys. Rev. Lett.}\ }\textbf {\bibinfo {volume} {106}},\ \bibinfo {pages}
  {112501} (\bibinfo {year} {2011}{\natexlab{a}})}\BibitemShut {NoStop}%
\bibitem [{\citenamefont {{Zhang Y. H.}}\ \emph {et~al.}(2016)\citenamefont
  {{Zhang Y. H.}}, \citenamefont {Litvinov}, \citenamefont {Uesaka},\ and\
  \citenamefont {Xu}}]{12}%
  \BibitemOpen
  \bibfield  {author} {\bibinfo {author} {\bibnamefont {{Zhang Y. H.}}},
  \bibinfo {author} {\bibfnamefont {Y.~A.}\ \bibnamefont {Litvinov}}, \bibinfo
  {author} {\bibfnamefont {T.}~\bibnamefont {Uesaka}}, \ and\ \bibinfo {author}
  {\bibfnamefont {H.~S.}\ \bibnamefont {Xu}},\ }\href {\doibase
  10.1088/0031-8949/91/7/073002} {\bibfield  {journal} {\bibinfo  {journal}
  {Phys. Scr.}\ }\textbf {\bibinfo {volume} {91}},\ \bibinfo {pages} {073002}
  (\bibinfo {year} {2016})}\BibitemShut {NoStop}%
\bibitem [{\citenamefont {{Zhang W.}}\ \emph {et~al.}(2014)\citenamefont
  {{Zhang W.}}, \citenamefont {Tu}, \citenamefont {Wang}, \citenamefont
  {Zhang}, \citenamefont {Xu}, \citenamefont {Litvinov}, \citenamefont {Blaum},
  \citenamefont {Chen}, \citenamefont {Chen}, \citenamefont {Fu}, \citenamefont
  {Ge}, \citenamefont {Gao}, \citenamefont {Hu}, \citenamefont {Huang},
  \citenamefont {Litvinov}, \citenamefont {Liu}, \citenamefont {Ma},
  \citenamefont {Mao}, \citenamefont {Mei}, \citenamefont {Shuai},
  \citenamefont {Sun}, \citenamefont {Xia}, \citenamefont {Xiao}, \citenamefont
  {Xing}, \citenamefont {Xu}, \citenamefont {Yamaguchi}, \citenamefont {Yan},
  \citenamefont {Yang}, \citenamefont {Yuan}, \citenamefont {Zeng},
  \citenamefont {Zhang}, \citenamefont {Zhao}, \citenamefont {Zhao},\ and\
  \citenamefont {Zhou}}]{15}%
  \BibitemOpen
  \bibfield  {author} {\bibinfo {author} {\bibnamefont {{Zhang W.}}}, \bibinfo
  {author} {\bibfnamefont {X.}~\bibnamefont {Tu}}, \bibinfo {author}
  {\bibfnamefont {M.}~\bibnamefont {Wang}}, \bibinfo {author} {\bibfnamefont
  {Y.}~\bibnamefont {Zhang}}, \bibinfo {author} {\bibfnamefont
  {H.}~\bibnamefont {Xu}}, \bibinfo {author} {\bibfnamefont {Y.~A.}\
  \bibnamefont {Litvinov}}, \bibinfo {author} {\bibfnamefont {K.}~\bibnamefont
  {Blaum}}, \bibinfo {author} {\bibfnamefont {R.}~\bibnamefont {Chen}},
  \bibinfo {author} {\bibfnamefont {X.}~\bibnamefont {Chen}}, \bibinfo {author}
  {\bibfnamefont {C.}~\bibnamefont {Fu}}, \bibinfo {author} {\bibfnamefont
  {Z.}~\bibnamefont {Ge}}, \bibinfo {author} {\bibfnamefont {B.}~\bibnamefont
  {Gao}}, \bibinfo {author} {\bibfnamefont {Z.}~\bibnamefont {Hu}}, \bibinfo
  {author} {\bibfnamefont {W.}~\bibnamefont {Huang}}, \bibinfo {author}
  {\bibfnamefont {S.}~\bibnamefont {Litvinov}}, \bibinfo {author}
  {\bibfnamefont {D.}~\bibnamefont {Liu}}, \bibinfo {author} {\bibfnamefont
  {X.}~\bibnamefont {Ma}}, \bibinfo {author} {\bibfnamefont {R.}~\bibnamefont
  {Mao}}, \bibinfo {author} {\bibfnamefont {B.}~\bibnamefont {Mei}}, \bibinfo
  {author} {\bibfnamefont {P.}~\bibnamefont {Shuai}}, \bibinfo {author}
  {\bibfnamefont {B.}~\bibnamefont {Sun}}, \bibinfo {author} {\bibfnamefont
  {J.}~\bibnamefont {Xia}}, \bibinfo {author} {\bibfnamefont {G.}~\bibnamefont
  {Xiao}}, \bibinfo {author} {\bibfnamefont {Y.}~\bibnamefont {Xing}}, \bibinfo
  {author} {\bibfnamefont {X.}~\bibnamefont {Xu}}, \bibinfo {author}
  {\bibfnamefont {T.}~\bibnamefont {Yamaguchi}}, \bibinfo {author}
  {\bibfnamefont {X.}~\bibnamefont {Yan}}, \bibinfo {author} {\bibfnamefont
  {J.}~\bibnamefont {Yang}}, \bibinfo {author} {\bibfnamefont {Y.}~\bibnamefont
  {Yuan}}, \bibinfo {author} {\bibfnamefont {Q.}~\bibnamefont {Zeng}}, \bibinfo
  {author} {\bibfnamefont {X.}~\bibnamefont {Zhang}}, \bibinfo {author}
  {\bibfnamefont {H.}~\bibnamefont {Zhao}}, \bibinfo {author} {\bibfnamefont
  {T.}~\bibnamefont {Zhao}}, \ and\ \bibinfo {author} {\bibfnamefont
  {X.}~\bibnamefont {Zhou}},\ }\href {\doibase 10.1016/j.nima.2014.04.051}
  {\bibfield  {journal} {\bibinfo  {journal} {Nucl. Instrum. Methods Phys.
  Res., Sect. A}\ }\textbf {\bibinfo {volume} {756}},\ \bibinfo {pages} {1}
  (\bibinfo {year} {2014})}\BibitemShut {NoStop}%
\bibitem [{\citenamefont {Yan}\ \emph {et~al.}(2019)\citenamefont {Yan},
  \citenamefont {Chen}, \citenamefont {Wang}, \citenamefont {Yuan},
  \citenamefont {Yuan}, \citenamefont {Wang}, \citenamefont {Cai},
  \citenamefont {Zhang}, \citenamefont {Lu}, \citenamefont {Fu}, \citenamefont
  {Zhou}, \citenamefont {Zhao}, \citenamefont {Litvinov},\ and\ \citenamefont
  {Zhang}}]{14}%
  \BibitemOpen
  \bibfield  {author} {\bibinfo {author} {\bibfnamefont {X.-L.}\ \bibnamefont
  {Yan}}, \bibinfo {author} {\bibfnamefont {R.-J.}\ \bibnamefont {Chen}},
  \bibinfo {author} {\bibfnamefont {M.}~\bibnamefont {Wang}}, \bibinfo {author}
  {\bibfnamefont {Y.-J.}\ \bibnamefont {Yuan}}, \bibinfo {author}
  {\bibfnamefont {J.-D.}\ \bibnamefont {Yuan}}, \bibinfo {author}
  {\bibfnamefont {S.-M.}\ \bibnamefont {Wang}}, \bibinfo {author}
  {\bibfnamefont {G.-Z.}\ \bibnamefont {Cai}}, \bibinfo {author} {\bibfnamefont
  {M.}~\bibnamefont {Zhang}}, \bibinfo {author} {\bibfnamefont {Z.-W.}\
  \bibnamefont {Lu}}, \bibinfo {author} {\bibfnamefont {C.-Y.}\ \bibnamefont
  {Fu}}, \bibinfo {author} {\bibfnamefont {X.}~\bibnamefont {Zhou}}, \bibinfo
  {author} {\bibfnamefont {D.-M.}\ \bibnamefont {Zhao}}, \bibinfo {author}
  {\bibfnamefont {Y.~A.}\ \bibnamefont {Litvinov}}, \ and\ \bibinfo {author}
  {\bibfnamefont {Y.-H.}\ \bibnamefont {Zhang}},\ }\href {\doibase
  10.1016/j.nima.2019.03.058} {\bibfield  {journal} {\bibinfo  {journal} {Nucl.
  Instrum. Methods Phys. Res., Sect. A}\ }\textbf {\bibinfo {volume} {931}},\
  \bibinfo {pages} {52} (\bibinfo {year} {2019})}\BibitemShut {NoStop}%
\bibitem [{\citenamefont {Zhou}\ \emph {et~al.}(2021)\citenamefont {Zhou},
  \citenamefont {Zhang}, \citenamefont {Wang}, \citenamefont {Zhang},
  \citenamefont {Yuan}, \citenamefont {Yan}, \citenamefont {Zhou},
  \citenamefont {Xu}, \citenamefont {Chen}, \citenamefont {Xing}, \citenamefont
  {Chen}, \citenamefont {Xu}, \citenamefont {Shuai}, \citenamefont {Fu},
  \citenamefont {Zeng}, \citenamefont {Sun}, \citenamefont {Li}, \citenamefont
  {Wang}, \citenamefont {Bao}, \citenamefont {Si}, \citenamefont {Deng},
  \citenamefont {Liu}, \citenamefont {Liao}, \citenamefont {Shi}, \citenamefont
  {Song}, \citenamefont {Yang}, \citenamefont {Ge}, \citenamefont {Litvinov},
  \citenamefont {Litvinov}, \citenamefont {Sidhu}, \citenamefont {Yamaguchi},
  \citenamefont {Omika}, \citenamefont {Wakayama}, \citenamefont {Suzuki},\
  and\ \citenamefont {Moriguchi}}]{16}%
  \BibitemOpen
  \bibfield  {author} {\bibinfo {author} {\bibfnamefont {X.}~\bibnamefont
  {Zhou}}, \bibinfo {author} {\bibfnamefont {M.}~\bibnamefont {Zhang}},
  \bibinfo {author} {\bibfnamefont {M.}~\bibnamefont {Wang}}, \bibinfo {author}
  {\bibfnamefont {Y.~H.}\ \bibnamefont {Zhang}}, \bibinfo {author}
  {\bibfnamefont {Y.~J.}\ \bibnamefont {Yuan}}, \bibinfo {author}
  {\bibfnamefont {X.~L.}\ \bibnamefont {Yan}}, \bibinfo {author} {\bibfnamefont
  {X.~H.}\ \bibnamefont {Zhou}}, \bibinfo {author} {\bibfnamefont {H.~S.}\
  \bibnamefont {Xu}}, \bibinfo {author} {\bibfnamefont {X.~C.}\ \bibnamefont
  {Chen}}, \bibinfo {author} {\bibfnamefont {Y.~M.}\ \bibnamefont {Xing}},
  \bibinfo {author} {\bibfnamefont {R.~J.}\ \bibnamefont {Chen}}, \bibinfo
  {author} {\bibfnamefont {X.}~\bibnamefont {Xu}}, \bibinfo {author}
  {\bibfnamefont {P.}~\bibnamefont {Shuai}}, \bibinfo {author} {\bibfnamefont
  {C.~Y.}\ \bibnamefont {Fu}}, \bibinfo {author} {\bibfnamefont
  {Q.}~\bibnamefont {Zeng}}, \bibinfo {author} {\bibfnamefont {M.~Z.}\
  \bibnamefont {Sun}}, \bibinfo {author} {\bibfnamefont {H.~F.}\ \bibnamefont
  {Li}}, \bibinfo {author} {\bibfnamefont {Q.}~\bibnamefont {Wang}}, \bibinfo
  {author} {\bibfnamefont {T.}~\bibnamefont {Bao}}, \bibinfo {author}
  {\bibfnamefont {M.}~\bibnamefont {Si}}, \bibinfo {author} {\bibfnamefont
  {H.~Y.}\ \bibnamefont {Deng}}, \bibinfo {author} {\bibfnamefont {M.~Z.}\
  \bibnamefont {Liu}}, \bibinfo {author} {\bibfnamefont {T.}~\bibnamefont
  {Liao}}, \bibinfo {author} {\bibfnamefont {J.~Y.}\ \bibnamefont {Shi}},
  \bibinfo {author} {\bibfnamefont {Y.~N.}\ \bibnamefont {Song}}, \bibinfo
  {author} {\bibfnamefont {J.~C.}\ \bibnamefont {Yang}}, \bibinfo {author}
  {\bibfnamefont {W.~W.}\ \bibnamefont {Ge}}, \bibinfo {author} {\bibfnamefont
  {Y.~A.}\ \bibnamefont {Litvinov}}, \bibinfo {author} {\bibfnamefont {S.~A.}\
  \bibnamefont {Litvinov}}, \bibinfo {author} {\bibfnamefont {R.~S.}\
  \bibnamefont {Sidhu}}, \bibinfo {author} {\bibfnamefont {T.}~\bibnamefont
  {Yamaguchi}}, \bibinfo {author} {\bibfnamefont {S.}~\bibnamefont {Omika}},
  \bibinfo {author} {\bibfnamefont {K.}~\bibnamefont {Wakayama}}, \bibinfo
  {author} {\bibfnamefont {S.}~\bibnamefont {Suzuki}}, \ and\ \bibinfo {author}
  {\bibfnamefont {T.}~\bibnamefont {Moriguchi}},\ }\href {\doibase
  10.1103/PhysRevAccelBeams.24.042802} {\bibfield  {journal} {\bibinfo
  {journal} {Phys. Rev. Accel. Beams}\ }\textbf {\bibinfo {volume} {24}},\
  \bibinfo {pages} {042802} (\bibinfo {year} {2021})}\BibitemShut {NoStop}%
\bibitem [{\citenamefont {Tu}\ \emph {et~al.}(2011{\natexlab{b}})\citenamefont
  {Tu}, \citenamefont {Wang}, \citenamefont {Litvinov}, \citenamefont {Zhang},
  \citenamefont {Xu}, \citenamefont {Sun}, \citenamefont {Audi}, \citenamefont
  {Blaum}, \citenamefont {Du}, \citenamefont {Huang}, \citenamefont {Hu},
  \citenamefont {Geng}, \citenamefont {Jin}, \citenamefont {Liu}, \citenamefont
  {Liu}, \citenamefont {Mei}, \citenamefont {Mao}, \citenamefont {Ma},
  \citenamefont {Suzuki}, \citenamefont {Shuai}, \citenamefont {Sun},
  \citenamefont {Tang}, \citenamefont {Wang}, \citenamefont {Wang},
  \citenamefont {Xiao}, \citenamefont {Xu}, \citenamefont {Xia}, \citenamefont
  {Yang}, \citenamefont {Ye}, \citenamefont {Yamaguchi}, \citenamefont {Yan},
  \citenamefont {Yuan}, \citenamefont {Yamaguchi}, \citenamefont {Zang},
  \citenamefont {Zhao}, \citenamefont {Zhao}, \citenamefont {Zhang},
  \citenamefont {Zhou},\ and\ \citenamefont {Zhan}}]{tu_nima_2011}%
  \BibitemOpen
  \bibfield  {author} {\bibinfo {author} {\bibfnamefont {X.~L.}\ \bibnamefont
  {Tu}}, \bibinfo {author} {\bibfnamefont {M.}~\bibnamefont {Wang}}, \bibinfo
  {author} {\bibfnamefont {Y.~A.}\ \bibnamefont {Litvinov}}, \bibinfo {author}
  {\bibfnamefont {Y.~H.}\ \bibnamefont {Zhang}}, \bibinfo {author}
  {\bibfnamefont {H.~S.}\ \bibnamefont {Xu}}, \bibinfo {author} {\bibfnamefont
  {Z.~Y.}\ \bibnamefont {Sun}}, \bibinfo {author} {\bibfnamefont
  {G.}~\bibnamefont {Audi}}, \bibinfo {author} {\bibfnamefont {K.}~\bibnamefont
  {Blaum}}, \bibinfo {author} {\bibfnamefont {C.~M.}\ \bibnamefont {Du}},
  \bibinfo {author} {\bibfnamefont {W.~X.}\ \bibnamefont {Huang}}, \bibinfo
  {author} {\bibfnamefont {Z.~G.}\ \bibnamefont {Hu}}, \bibinfo {author}
  {\bibfnamefont {P.}~\bibnamefont {Geng}}, \bibinfo {author} {\bibfnamefont
  {S.~L.}\ \bibnamefont {Jin}}, \bibinfo {author} {\bibfnamefont {L.~X.}\
  \bibnamefont {Liu}}, \bibinfo {author} {\bibfnamefont {Y.}~\bibnamefont
  {Liu}}, \bibinfo {author} {\bibfnamefont {B.}~\bibnamefont {Mei}}, \bibinfo
  {author} {\bibfnamefont {R.~S.}\ \bibnamefont {Mao}}, \bibinfo {author}
  {\bibfnamefont {X.~W.}\ \bibnamefont {Ma}}, \bibinfo {author} {\bibfnamefont
  {H.}~\bibnamefont {Suzuki}}, \bibinfo {author} {\bibfnamefont
  {P.}~\bibnamefont {Shuai}}, \bibinfo {author} {\bibfnamefont
  {Y.}~\bibnamefont {Sun}}, \bibinfo {author} {\bibfnamefont {S.~W.}\
  \bibnamefont {Tang}}, \bibinfo {author} {\bibfnamefont {J.~S.}\ \bibnamefont
  {Wang}}, \bibinfo {author} {\bibfnamefont {S.~T.}\ \bibnamefont {Wang}},
  \bibinfo {author} {\bibfnamefont {G.~Q.}\ \bibnamefont {Xiao}}, \bibinfo
  {author} {\bibfnamefont {X.}~\bibnamefont {Xu}}, \bibinfo {author}
  {\bibfnamefont {J.~W.}\ \bibnamefont {Xia}}, \bibinfo {author} {\bibfnamefont
  {J.~C.}\ \bibnamefont {Yang}}, \bibinfo {author} {\bibfnamefont {R.~P.}\
  \bibnamefont {Ye}}, \bibinfo {author} {\bibfnamefont {T.}~\bibnamefont
  {Yamaguchi}}, \bibinfo {author} {\bibfnamefont {X.~L.}\ \bibnamefont {Yan}},
  \bibinfo {author} {\bibfnamefont {Y.~J.}\ \bibnamefont {Yuan}}, \bibinfo
  {author} {\bibfnamefont {Y.}~\bibnamefont {Yamaguchi}}, \bibinfo {author}
  {\bibfnamefont {Y.~D.}\ \bibnamefont {Zang}}, \bibinfo {author}
  {\bibfnamefont {H.~W.}\ \bibnamefont {Zhao}}, \bibinfo {author}
  {\bibfnamefont {T.~C.}\ \bibnamefont {Zhao}}, \bibinfo {author}
  {\bibfnamefont {X.~Y.}\ \bibnamefont {Zhang}}, \bibinfo {author}
  {\bibfnamefont {X.~H.}\ \bibnamefont {Zhou}}, \ and\ \bibinfo {author}
  {\bibfnamefont {W.~L.}\ \bibnamefont {Zhan}},\ }\href {\doibase
  10.1016/j.nima.2011.07.018} {\bibfield  {journal} {\bibinfo  {journal}
  {Nuclear Instruments and Methods in Physics Research Section A: Accelerators,
  Spectrometers, Detectors and Associated Equipment}\ }\textbf {\bibinfo
  {volume} {654}},\ \bibinfo {pages} {213} (\bibinfo {year}
  {2011}{\natexlab{b}})}\BibitemShut {NoStop}%
\bibitem [{\citenamefont {Kondev}\ \emph {et~al.}(2021)\citenamefont {Kondev},
  \citenamefont {Wang}, \citenamefont {Huang}, \citenamefont {Naimi},\ and\
  \citenamefont {G}}]{kondev_nubase2020_2021}%
  \BibitemOpen
  \bibfield  {author} {\bibinfo {author} {\bibfnamefont {F.~G.}\ \bibnamefont
  {Kondev}}, \bibinfo {author} {\bibfnamefont {M.}~\bibnamefont {Wang}},
  \bibinfo {author} {\bibfnamefont {W.~J.}\ \bibnamefont {Huang}}, \bibinfo
  {author} {\bibfnamefont {S.}~\bibnamefont {Naimi}}, \ and\ \bibinfo {author}
  {\bibfnamefont {A.}~\bibnamefont {G}},\ }\href {\doibase
  10.1088/1674-1137/abddae} {\bibfield  {journal} {\bibinfo  {journal} {Chin.
  Phys. C}\ }\textbf {\bibinfo {volume} {45}},\ \bibinfo {pages} {030001}
  (\bibinfo {year} {2021})}\BibitemShut {NoStop}%
\bibitem [{\citenamefont {Cheng}\ \emph {et~al.}(2017)\citenamefont {Cheng},
  \citenamefont {Jiang}, \citenamefont {Zhao},\ and\ \citenamefont
  {Arima}}]{24}%
  \BibitemOpen
  \bibfield  {author} {\bibinfo {author} {\bibfnamefont {Y.~Y.}\ \bibnamefont
  {Cheng}}, \bibinfo {author} {\bibfnamefont {H.}~\bibnamefont {Jiang}},
  \bibinfo {author} {\bibfnamefont {Y.~M.}\ \bibnamefont {Zhao}}, \ and\
  \bibinfo {author} {\bibfnamefont {A.}~\bibnamefont {Arima}},\ }\href
  {\doibase 10.1088/1361-6471/aa8a25} {\bibfield  {journal} {\bibinfo
  {journal} {J. Phys. G: Nucl. Part. Phys.}\ }\textbf {\bibinfo {volume}
  {44}},\ \bibinfo {pages} {115102} (\bibinfo {year} {2017})}\BibitemShut
  {NoStop}%
\bibitem [{\citenamefont {Zong}\ \emph {et~al.}(2022)\citenamefont {Zong},
  \citenamefont {Ma}, \citenamefont {Lin},\ and\ \citenamefont {Zhao}}]{25}%
  \BibitemOpen
  \bibfield  {author} {\bibinfo {author} {\bibfnamefont {Y.~Y.}\ \bibnamefont
  {Zong}}, \bibinfo {author} {\bibfnamefont {C.}~\bibnamefont {Ma}}, \bibinfo
  {author} {\bibfnamefont {M.~Q.}\ \bibnamefont {Lin}}, \ and\ \bibinfo
  {author} {\bibfnamefont {Y.~M.}\ \bibnamefont {Zhao}},\ }\href {\doibase
  10.1103/PhysRevC.105.034321} {\bibfield  {journal} {\bibinfo  {journal}
  {Phys. Rev. C}\ }\textbf {\bibinfo {volume} {105}},\ \bibinfo {pages}
  {034321} (\bibinfo {year} {2022})}\BibitemShut {NoStop}%
\bibitem [{\citenamefont {Tian}\ \emph {et~al.}(2013)\citenamefont {Tian},
  \citenamefont {Wang}, \citenamefont {Li},\ and\ \citenamefont {Li}}]{26}%
  \BibitemOpen
  \bibfield  {author} {\bibinfo {author} {\bibfnamefont {J.}~\bibnamefont
  {Tian}}, \bibinfo {author} {\bibfnamefont {N.}~\bibnamefont {Wang}}, \bibinfo
  {author} {\bibfnamefont {C.}~\bibnamefont {Li}}, \ and\ \bibinfo {author}
  {\bibfnamefont {J.}~\bibnamefont {Li}},\ }\href {\doibase
  10.1103/PhysRevC.87.014313} {\bibfield  {journal} {\bibinfo  {journal} {Phys.
  Rev. C}\ }\textbf {\bibinfo {volume} {87}},\ \bibinfo {pages} {014313}
  (\bibinfo {year} {2013})}\BibitemShut {NoStop}%
\bibitem [{\citenamefont {Yuan}\ \emph {et~al.}(2014)\citenamefont {Yuan},
  \citenamefont {Qi}, \citenamefont {Xu}, \citenamefont {Suzuki},\ and\
  \citenamefont {Otsuka}}]{27}%
  \BibitemOpen
  \bibfield  {author} {\bibinfo {author} {\bibfnamefont {C.}~\bibnamefont
  {Yuan}}, \bibinfo {author} {\bibfnamefont {C.}~\bibnamefont {Qi}}, \bibinfo
  {author} {\bibfnamefont {F.}~\bibnamefont {Xu}}, \bibinfo {author}
  {\bibfnamefont {T.}~\bibnamefont {Suzuki}}, \ and\ \bibinfo {author}
  {\bibfnamefont {T.}~\bibnamefont {Otsuka}},\ }\href {\doibase
  10.1103/PhysRevC.89.044327} {\bibfield  {journal} {\bibinfo  {journal} {Phys.
  Rev. C}\ }\textbf {\bibinfo {volume} {89}},\ \bibinfo {pages} {044327}
  (\bibinfo {year} {2014})}\BibitemShut {NoStop}%
\bibitem [{\citenamefont {Fortune}(2018)}]{28}%
  \BibitemOpen
  \bibfield  {author} {\bibinfo {author} {\bibfnamefont {H.~T.}\ \bibnamefont
  {Fortune}},\ }\href {\doibase 10.1103/PhysRevC.97.034301} {\bibfield
  {journal} {\bibinfo  {journal} {Phys. Rev. C}\ }\textbf {\bibinfo {volume}
  {97}},\ \bibinfo {pages} {034301} (\bibinfo {year} {2018})}\BibitemShut
  {NoStop}%
\bibitem [{\citenamefont {MacCormick}\ and\ \citenamefont
  {Audi}(2014)}]{MACCORMICK201461}%
  \BibitemOpen
  \bibfield  {author} {\bibinfo {author} {\bibfnamefont {M.}~\bibnamefont
  {MacCormick}}\ and\ \bibinfo {author} {\bibfnamefont {G.}~\bibnamefont
  {Audi}},\ }\href {\doibase https://doi.org/10.1016/j.nuclphysa.2014.01.007}
  {\bibfield  {journal} {\bibinfo  {journal} {Nucl. Phys. A}\ }\textbf
  {\bibinfo {volume} {925}},\ \bibinfo {pages} {61} (\bibinfo {year}
  {2014})}\BibitemShut {NoStop}%
\bibitem [{\citenamefont {Bentley}\ and\ \citenamefont
  {Lenzi}(2007)}]{bentley_coulomb_2007}%
  \BibitemOpen
  \bibfield  {author} {\bibinfo {author} {\bibfnamefont {M.~A.}\ \bibnamefont
  {Bentley}}\ and\ \bibinfo {author} {\bibfnamefont {S.~M.}\ \bibnamefont
  {Lenzi}},\ }\href {\doibase https://doi.org/10.1016/j.ppnp.2006.10.001}
  {\bibfield  {journal} {\bibinfo  {journal} {Prog. Part. Nucl. Phys.}\
  }\textbf {\bibinfo {volume} {59}},\ \bibinfo {pages} {497} (\bibinfo {year}
  {2007})}\BibitemShut {NoStop}%
\bibitem [{\citenamefont {Boso}\ \emph {et~al.}(2018)\citenamefont {Boso},
  \citenamefont {Lenzi}, \citenamefont {Recchia}, \citenamefont {Bonnard},
  \citenamefont {Zuker}, \citenamefont {Aydin}, \citenamefont {Bentley},
  \citenamefont {Cederwall}, \citenamefont {Clement}, \citenamefont
  {de~France}, \citenamefont {Di~Nitto}, \citenamefont {Dijon}, \citenamefont
  {Doncel}, \citenamefont {Ghazi-Moradi}, \citenamefont {Gadea}, \citenamefont
  {Gottardo}, \citenamefont {Henry}, \citenamefont {Hüyük}, \citenamefont
  {Jaworski}, \citenamefont {John}, \citenamefont {Juhász}, \citenamefont
  {Kuti}, \citenamefont {Melon}, \citenamefont {Mengoni}, \citenamefont
  {Michelagnoli}, \citenamefont {Modamio}, \citenamefont {Napoli},
  \citenamefont {Nyakó}, \citenamefont {Nyberg}, \citenamefont {Palacz},
  \citenamefont {Timár},\ and\ \citenamefont
  {Valiente-Dobón}}]{boso_neutron_2018}%
  \BibitemOpen
  \bibfield  {author} {\bibinfo {author} {\bibfnamefont {A.}~\bibnamefont
  {Boso}}, \bibinfo {author} {\bibfnamefont {S.~M.}\ \bibnamefont {Lenzi}},
  \bibinfo {author} {\bibfnamefont {F.}~\bibnamefont {Recchia}}, \bibinfo
  {author} {\bibfnamefont {J.}~\bibnamefont {Bonnard}}, \bibinfo {author}
  {\bibfnamefont {A.~P.}\ \bibnamefont {Zuker}}, \bibinfo {author}
  {\bibfnamefont {S.}~\bibnamefont {Aydin}}, \bibinfo {author} {\bibfnamefont
  {M.~A.}\ \bibnamefont {Bentley}}, \bibinfo {author} {\bibfnamefont
  {B.}~\bibnamefont {Cederwall}}, \bibinfo {author} {\bibfnamefont
  {E.}~\bibnamefont {Clement}}, \bibinfo {author} {\bibfnamefont
  {G.}~\bibnamefont {de~France}}, \bibinfo {author} {\bibfnamefont
  {A.}~\bibnamefont {Di~Nitto}}, \bibinfo {author} {\bibfnamefont
  {A.}~\bibnamefont {Dijon}}, \bibinfo {author} {\bibfnamefont
  {M.}~\bibnamefont {Doncel}}, \bibinfo {author} {\bibfnamefont
  {F.}~\bibnamefont {Ghazi-Moradi}}, \bibinfo {author} {\bibfnamefont
  {A.}~\bibnamefont {Gadea}}, \bibinfo {author} {\bibfnamefont
  {A.}~\bibnamefont {Gottardo}}, \bibinfo {author} {\bibfnamefont
  {T.}~\bibnamefont {Henry}}, \bibinfo {author} {\bibfnamefont
  {T.}~\bibnamefont {Hüyük}}, \bibinfo {author} {\bibfnamefont
  {G.}~\bibnamefont {Jaworski}}, \bibinfo {author} {\bibfnamefont {P.~R.}\
  \bibnamefont {John}}, \bibinfo {author} {\bibfnamefont {K.}~\bibnamefont
  {Juhász}}, \bibinfo {author} {\bibfnamefont {I.}~\bibnamefont {Kuti}},
  \bibinfo {author} {\bibfnamefont {B.}~\bibnamefont {Melon}}, \bibinfo
  {author} {\bibfnamefont {D.}~\bibnamefont {Mengoni}}, \bibinfo {author}
  {\bibfnamefont {C.}~\bibnamefont {Michelagnoli}}, \bibinfo {author}
  {\bibfnamefont {V.}~\bibnamefont {Modamio}}, \bibinfo {author} {\bibfnamefont
  {D.~R.}\ \bibnamefont {Napoli}}, \bibinfo {author} {\bibfnamefont {B.~M.}\
  \bibnamefont {Nyakó}}, \bibinfo {author} {\bibfnamefont {J.}~\bibnamefont
  {Nyberg}}, \bibinfo {author} {\bibfnamefont {M.}~\bibnamefont {Palacz}},
  \bibinfo {author} {\bibfnamefont {J.}~\bibnamefont {Timár}}, \ and\ \bibinfo
  {author} {\bibfnamefont {J.~J.}\ \bibnamefont {Valiente-Dobón}},\ }\href
  {\doibase 10.1103/PhysRevLett.121.032502} {\bibfield  {journal} {\bibinfo
  {journal} {Phys. Rev. Lett.}\ }\textbf {\bibinfo {volume} {121}},\ \bibinfo
  {pages} {032502} (\bibinfo {year} {2018})}\BibitemShut {NoStop}%
\bibitem [{ims()}]{imsrg_code}%
  \BibitemOpen
  \href@noop {} {}\bibinfo {howpublished}
  {\url{https://github.com/ragnarstroberg/imsrg}}\BibitemShut {NoStop}%
\bibitem [{\citenamefont {Shimizu}\ \emph {et~al.}(2019)\citenamefont
  {Shimizu}, \citenamefont {Mizusaki}, \citenamefont {Utsuno},\ and\
  \citenamefont {Tsunoda}}]{39}%
  \BibitemOpen
  \bibfield  {author} {\bibinfo {author} {\bibfnamefont {N.}~\bibnamefont
  {Shimizu}}, \bibinfo {author} {\bibfnamefont {T.}~\bibnamefont {Mizusaki}},
  \bibinfo {author} {\bibfnamefont {Y.}~\bibnamefont {Utsuno}}, \ and\ \bibinfo
  {author} {\bibfnamefont {Y.}~\bibnamefont {Tsunoda}},\ }\href {\doibase
  https://doi.org/10.1016/j.cpc.2019.06.011} {\bibfield  {journal} {\bibinfo
  {journal} {Comput. Phys. Commun.}\ }\textbf {\bibinfo {volume} {244}},\
  \bibinfo {pages} {372} (\bibinfo {year} {2019})}\BibitemShut {NoStop}%
\bibitem [{\citenamefont {Entem}\ and\ \citenamefont
  {Machleidt}(2003)}]{PhysRevC.68.041001}%
  \BibitemOpen
  \bibfield  {author} {\bibinfo {author} {\bibfnamefont {D.~R.}\ \bibnamefont
  {Entem}}\ and\ \bibinfo {author} {\bibfnamefont {R.}~\bibnamefont
  {Machleidt}},\ }\href {\doibase 10.1103/PhysRevC.68.041001} {\bibfield
  {journal} {\bibinfo  {journal} {Phys. Rev. C}\ }\textbf {\bibinfo {volume}
  {68}},\ \bibinfo {pages} {041001(R)} (\bibinfo {year} {2003})}\BibitemShut
  {NoStop}%
\bibitem [{\citenamefont {Bogner}\ \emph {et~al.}(2007)\citenamefont {Bogner},
  \citenamefont {Furnstahl},\ and\ \citenamefont {Perry}}]{PhysRevC.75.061001}%
  \BibitemOpen
  \bibfield  {author} {\bibinfo {author} {\bibfnamefont {S.~K.}\ \bibnamefont
  {Bogner}}, \bibinfo {author} {\bibfnamefont {R.~J.}\ \bibnamefont
  {Furnstahl}}, \ and\ \bibinfo {author} {\bibfnamefont {R.~J.}\ \bibnamefont
  {Perry}},\ }\href {\doibase 10.1103/PhysRevC.75.061001} {\bibfield  {journal}
  {\bibinfo  {journal} {Phys. Rev. C}\ }\textbf {\bibinfo {volume} {75}},\
  \bibinfo {pages} {061001} (\bibinfo {year} {2007})}\BibitemShut {NoStop}%
\bibitem [{\citenamefont {Miyagi}\ \emph {et~al.}(2022)\citenamefont {Miyagi},
  \citenamefont {Stroberg}, \citenamefont {Navr\'atil}, \citenamefont
  {Hebeler},\ and\ \citenamefont {Holt}}]{PhysRevC.105.014302}%
  \BibitemOpen
  \bibfield  {author} {\bibinfo {author} {\bibfnamefont {T.}~\bibnamefont
  {Miyagi}}, \bibinfo {author} {\bibfnamefont {S.~R.}\ \bibnamefont
  {Stroberg}}, \bibinfo {author} {\bibfnamefont {P.}~\bibnamefont
  {Navr\'atil}}, \bibinfo {author} {\bibfnamefont {K.}~\bibnamefont {Hebeler}},
  \ and\ \bibinfo {author} {\bibfnamefont {J.~D.}\ \bibnamefont {Holt}},\
  }\href {\doibase 10.1103/PhysRevC.105.014302} {\bibfield  {journal} {\bibinfo
   {journal} {Phys. Rev. C}\ }\textbf {\bibinfo {volume} {105}},\ \bibinfo
  {pages} {014302} (\bibinfo {year} {2022})}\BibitemShut {NoStop}%
\bibitem [{\citenamefont {Som\`a}\ \emph {et~al.}(2020)\citenamefont {Som\`a},
  \citenamefont {Navr\'atil}, \citenamefont {Raimondi}, \citenamefont
  {Barbieri},\ and\ \citenamefont {Duguet}}]{PhysRevC.101.014318}%
  \BibitemOpen
  \bibfield  {author} {\bibinfo {author} {\bibfnamefont {V.}~\bibnamefont
  {Som\`a}}, \bibinfo {author} {\bibfnamefont {P.}~\bibnamefont {Navr\'atil}},
  \bibinfo {author} {\bibfnamefont {F.}~\bibnamefont {Raimondi}}, \bibinfo
  {author} {\bibfnamefont {C.}~\bibnamefont {Barbieri}}, \ and\ \bibinfo
  {author} {\bibfnamefont {T.}~\bibnamefont {Duguet}},\ }\href {\doibase
  10.1103/PhysRevC.101.014318} {\bibfield  {journal} {\bibinfo  {journal}
  {Phys. Rev. C}\ }\textbf {\bibinfo {volume} {101}},\ \bibinfo {pages}
  {014318} (\bibinfo {year} {2020})}\BibitemShut {NoStop}%
\bibitem [{\citenamefont {Lapoux}\ \emph {et~al.}(2016)\citenamefont {Lapoux},
  \citenamefont {Somà}, \citenamefont {Barbieri}, \citenamefont {Hergert},
  \citenamefont {Holt},\ and\ \citenamefont {Stroberg}}]{34}%
  \BibitemOpen
  \bibfield  {author} {\bibinfo {author} {\bibfnamefont {V.}~\bibnamefont
  {Lapoux}}, \bibinfo {author} {\bibfnamefont {V.}~\bibnamefont {Somà}},
  \bibinfo {author} {\bibfnamefont {C.}~\bibnamefont {Barbieri}}, \bibinfo
  {author} {\bibfnamefont {H.}~\bibnamefont {Hergert}}, \bibinfo {author}
  {\bibfnamefont {J.~D.}\ \bibnamefont {Holt}}, \ and\ \bibinfo {author}
  {\bibfnamefont {S.~R.}\ \bibnamefont {Stroberg}},\ }\href {\doibase
  10.1103/PhysRevLett.117.052501} {\bibfield  {journal} {\bibinfo  {journal}
  {Phys. Rev. Lett.}\ }\textbf {\bibinfo {volume} {117}},\ \bibinfo {pages}
  {052501} (\bibinfo {year} {2016})}\BibitemShut {NoStop}%
\bibitem [{\citenamefont {Simonis}\ \emph {et~al.}(2017)\citenamefont
  {Simonis}, \citenamefont {Stroberg}, \citenamefont {Hebeler}, \citenamefont
  {Holt},\ and\ \citenamefont {Schwenk}}]{36}%
  \BibitemOpen
  \bibfield  {author} {\bibinfo {author} {\bibfnamefont {J.}~\bibnamefont
  {Simonis}}, \bibinfo {author} {\bibfnamefont {S.~R.}\ \bibnamefont
  {Stroberg}}, \bibinfo {author} {\bibfnamefont {K.}~\bibnamefont {Hebeler}},
  \bibinfo {author} {\bibfnamefont {J.~D.}\ \bibnamefont {Holt}}, \ and\
  \bibinfo {author} {\bibfnamefont {A.}~\bibnamefont {Schwenk}},\ }\href
  {\doibase 10.1103/PhysRevC.96.014303} {\bibfield  {journal} {\bibinfo
  {journal} {Phys. Rev. C}\ }\textbf {\bibinfo {volume} {96}},\ \bibinfo
  {pages} {014303} (\bibinfo {year} {2017})}\BibitemShut {NoStop}%
\bibitem [{\citenamefont {de~Groote}\ \emph {et~al.}(2020)\citenamefont
  {de~Groote}, \citenamefont {Billowes}, \citenamefont {Binnersley},
  \citenamefont {Bissell}, \citenamefont {Cocolios}, \citenamefont
  {Day~Goodacre}, \citenamefont {Farooq-Smith}, \citenamefont {Fedorov},
  \citenamefont {Flanagan}, \citenamefont {Franchoo}, \citenamefont
  {Garcia~Ruiz}, \citenamefont {Gins}, \citenamefont {Holt}, \citenamefont
  {Koszorus}, \citenamefont {Lynch}, \citenamefont {Miyagi}, \citenamefont
  {Nazarewicz}, \citenamefont {Neyens}, \citenamefont {Reinhard}, \citenamefont
  {Rothe}, \citenamefont {Stroke}, \citenamefont {Vernon}, \citenamefont
  {Wendt}, \citenamefont {Wilkins}, \citenamefont {Xu},\ and\ \citenamefont
  {Yang}}]{37}%
  \BibitemOpen
  \bibfield  {author} {\bibinfo {author} {\bibfnamefont {R.~P.}\ \bibnamefont
  {de~Groote}}, \bibinfo {author} {\bibfnamefont {J.}~\bibnamefont {Billowes}},
  \bibinfo {author} {\bibfnamefont {C.~L.}\ \bibnamefont {Binnersley}},
  \bibinfo {author} {\bibfnamefont {M.~L.}\ \bibnamefont {Bissell}}, \bibinfo
  {author} {\bibfnamefont {T.~E.}\ \bibnamefont {Cocolios}}, \bibinfo {author}
  {\bibfnamefont {T.}~\bibnamefont {Day~Goodacre}}, \bibinfo {author}
  {\bibfnamefont {G.~J.}\ \bibnamefont {Farooq-Smith}}, \bibinfo {author}
  {\bibfnamefont {D.~V.}\ \bibnamefont {Fedorov}}, \bibinfo {author}
  {\bibfnamefont {K.~T.}\ \bibnamefont {Flanagan}}, \bibinfo {author}
  {\bibfnamefont {S.}~\bibnamefont {Franchoo}}, \bibinfo {author}
  {\bibfnamefont {R.~F.}\ \bibnamefont {Garcia~Ruiz}}, \bibinfo {author}
  {\bibfnamefont {W.}~\bibnamefont {Gins}}, \bibinfo {author} {\bibfnamefont
  {J.~D.}\ \bibnamefont {Holt}}, \bibinfo {author} {\bibfnamefont
  {A.}~\bibnamefont {Koszorus}}, \bibinfo {author} {\bibfnamefont {K.~M.}\
  \bibnamefont {Lynch}}, \bibinfo {author} {\bibfnamefont {T.}~\bibnamefont
  {Miyagi}}, \bibinfo {author} {\bibfnamefont {W.}~\bibnamefont {Nazarewicz}},
  \bibinfo {author} {\bibfnamefont {G.}~\bibnamefont {Neyens}}, \bibinfo
  {author} {\bibfnamefont {P.-G.}\ \bibnamefont {Reinhard}}, \bibinfo {author}
  {\bibfnamefont {S.}~\bibnamefont {Rothe}}, \bibinfo {author} {\bibfnamefont
  {H.~H.}\ \bibnamefont {Stroke}}, \bibinfo {author} {\bibfnamefont {A.~R.}\
  \bibnamefont {Vernon}}, \bibinfo {author} {\bibfnamefont {K.~D.~A.}\
  \bibnamefont {Wendt}}, \bibinfo {author} {\bibfnamefont {S.~G.}\ \bibnamefont
  {Wilkins}}, \bibinfo {author} {\bibfnamefont {Z.~Y.}\ \bibnamefont {Xu}}, \
  and\ \bibinfo {author} {\bibfnamefont {X.~F.}\ \bibnamefont {Yang}},\ }\href
  {\doibase 10.1038/s41567-020-0868-y} {\bibfield  {journal} {\bibinfo
  {journal} {Nat. Phys.}\ }\textbf {\bibinfo {volume} {16}},\ \bibinfo {pages}
  {620} (\bibinfo {year} {2020})}\BibitemShut {NoStop}%
\bibitem [{\citenamefont {Ekstr\"om}\ \emph {et~al.}(2013)\citenamefont
  {Ekstr\"om}, \citenamefont {Baardsen}, \citenamefont {Forss\'en},
  \citenamefont {Hagen}, \citenamefont {Hjorth-Jensen}, \citenamefont {Jansen},
  \citenamefont {Machleidt}, \citenamefont {Nazarewicz}, \citenamefont
  {Papenbrock}, \citenamefont {Sarich},\ and\ \citenamefont
  {Wild}}]{PhysRevLett.110.192502}%
  \BibitemOpen
  \bibfield  {author} {\bibinfo {author} {\bibfnamefont {A.}~\bibnamefont
  {Ekstr\"om}}, \bibinfo {author} {\bibfnamefont {G.}~\bibnamefont {Baardsen}},
  \bibinfo {author} {\bibfnamefont {C.}~\bibnamefont {Forss\'en}}, \bibinfo
  {author} {\bibfnamefont {G.}~\bibnamefont {Hagen}}, \bibinfo {author}
  {\bibfnamefont {M.}~\bibnamefont {Hjorth-Jensen}}, \bibinfo {author}
  {\bibfnamefont {G.~R.}\ \bibnamefont {Jansen}}, \bibinfo {author}
  {\bibfnamefont {R.}~\bibnamefont {Machleidt}}, \bibinfo {author}
  {\bibfnamefont {W.}~\bibnamefont {Nazarewicz}}, \bibinfo {author}
  {\bibfnamefont {T.}~\bibnamefont {Papenbrock}}, \bibinfo {author}
  {\bibfnamefont {J.}~\bibnamefont {Sarich}}, \ and\ \bibinfo {author}
  {\bibfnamefont {S.~M.}\ \bibnamefont {Wild}},\ }\href {\doibase
  10.1103/PhysRevLett.110.192502} {\bibfield  {journal} {\bibinfo  {journal}
  {Phys. Rev. Lett.}\ }\textbf {\bibinfo {volume} {110}},\ \bibinfo {pages}
  {192502} (\bibinfo {year} {2013})}\BibitemShut {NoStop}%
\bibitem [{\citenamefont {Tanihata}\ \emph {et~al.}(2013)\citenamefont
  {Tanihata}, \citenamefont {Savajols},\ and\ \citenamefont
  {Kanungo}}]{tanihata_recent_2013}%
  \BibitemOpen
  \bibfield  {author} {\bibinfo {author} {\bibfnamefont {I.}~\bibnamefont
  {Tanihata}}, \bibinfo {author} {\bibfnamefont {H.}~\bibnamefont {Savajols}},
  \ and\ \bibinfo {author} {\bibfnamefont {R.}~\bibnamefont {Kanungo}},\ }\href
  {\doibase 10.1016/j.ppnp.2012.07.001} {\bibfield  {journal} {\bibinfo
  {journal} {Prog. Part. Nucl. Phys.}\ }\textbf {\bibinfo {volume} {68}},\
  \bibinfo {pages} {215} (\bibinfo {year} {2013})}\BibitemShut {NoStop}%
\end{thebibliography}%
\end{document}